\documentclass[journal]{IEEEtran}
\usepackage{ifpdf}
\usepackage{cite}
\usepackage{url}
\usepackage{amsmath}
\usepackage{multirow}
\usepackage{graphicx}
\usepackage{epstopdf}
\usepackage{booktabs}
\usepackage{longtable}
\usepackage{multicol}
\usepackage{amsfonts}

\begin{document}
\title{A New Small-World IoT Routing Mechanism\\ based on Cayley Graphs}

\author{Yuna Jiang, Xiaohu Ge,~\IEEEmembership{Senior Member,~IEEE,}
               Yi Zhong,~\IEEEmembership{Member,~IEEE,}
               Guoqiang Mao,~\IEEEmembership{Fellow,~IEEE,}
               and Yonghui Li,~\IEEEmembership{Fellow,~IEEE.}
\thanks{Yuna Jiang, Xiaohu Ge (Cooresponding author) and Yi Zhong are with School of Electronic Information and Communications, Huazhong University of Science
and Technology, Wuhan 430074, Hubei, P.R. China e-mail: xhge@mail.hust.edu.cn.}

\thanks{Guoqiang Mao is with the School of Electrical and Data Engineering, University of
Technology Sydney, Sydney, NSW 2007, Australia.}

\thanks{Yonghui Li is with the School
of Electrical and Information Engineering, The University of Sydney, Sydney, NSW 2000, Australia.}

\thanks{Copyright (c) 2019 IEEE. Personal use of this material is permitted. However, permission to use this material for any other purposes must be obtained from the IEEE by sending a request to pubs-permissions@ieee.org.}}

\markboth{}%
{Shell \MakeLowercase{\textit{et al.}}: }

\maketitle

\begin{abstract}
An increasing number of low-power Internet of Things (IoT) devices will be widely deployed in the near future. Considering the short-range communication of low-power devices, multi-hop transmissions will become an important transmission mechanism in IoT networks. It is crucial for low-power devices to transmit data over long distances via multi-hop in a low-delay and reliable way. Small-world characteristics of networks indicate that the network has an advantage of a small Average Shortest-path Length (ASL) and a high Average Clustering Coefficient (ACC). In this paper, a new IoT routing mechanism considering small-world characteristics is proposed to reduce the delay and improve the reliability. The ASL and ACC are derived for performance analysis of small-world characteristics in IoT networks based on Cayley graphs. Besides, the reliability and delay models are proposed for Small-World IoT based on Cayley grapHs (SWITCH). Simulation results demonstrate that SWITCH has lower delay and better reliability than that of conventional Nearest Neighboring Routing (NNR). Moreover, the maximum delay of SWITCH is reduced by 50.6\% compared with that by NNR.
\end{abstract}


\IEEEpeerreviewmaketitle

\section{INTRODUCTION}

\IEEEPARstart{A}{ccording} to Ericsson's forecast, there will be approximately 31.4 billion devices connected to the Internet by 2023 and 15.1 billion will be low-power devices for short-range communications \cite{b1}. Internet of Things (IoT) communications will have a wide range of applications, such as smart transportations, smart homes, smart grids, etc \cite{b2}\cite{b3}. A main challenge is to design the wireless networks to meet the requirements of the IoT traffic \cite{b4}\cite{b5}. It is crucial to provide a low end-to-end delay for many mission critical applications, such as telerobotic surgery, factory automation \cite{b6}. However, due to channel impairment, critical data may be delayed or lost during transmissions in IoT scenarios. These mission critical IoT communication systems have stringent Quality of Service (QoS) requirements and reliable data transmissions \cite{b7}\cite{b8}. It is difficult for low-power devices to transmit data to the destination in single hop because of the long transmission distance and channel attenuation\cite{b9}\cite{a1}. Therefore, multi-hop transmissions will become important transmission technique in IoT communication systems. A fundamental problem in wireless multi-hop networks is to determine whether the signals should be routed over many short hops or a smaller number of longer hops. Generally, the long hop transmission is suitable for delay-sensitive applications and the short hop transmission is preferred for applications with low transmission power \cite{b10}. In IoT networks, minimizing power consumption and achieving low end-to-end delay are both important to ensure successful end-to-end transmissions, which cannot be achieved via pure long hop network or pure short hop networks. Small-world characteristics indicate that a number of hops can be greatly reduced by adding some shortcuts into the networks. \cite{b11}. By exploring small-world characteristics, low power devices can transmit data through short hops for saving energy and at the same time the added shortcuts can reduce the total delay. Thus, it is crucial to design the IoT routing schemes harnessing small-world characteristics.

The network with small-world characteristics has an advantage of a small average shortest-path length (ASL) and a high average clustering coefficient (ACC) \cite{b12}. With the small ASL, the data can be transferred between two nodes with reduced delay and energy consumption. A large ACC helps the data to be spread more widely across the network \cite{b11}. The small-world  network model was first proposed by Watts and Strogatz and has later been known as Watts and Strogatz (WS) model. In this model, the network is first constructed by using the vertex ring and each edge is then rearranged with a certain probability which can be adjusted to control the conversion between the regular network and the small-world network \cite{b13}. However, the rearrangement of each edge in the WS model may damage the network connectivity. Newman and Watts proposed a new Newman and Watts (NW) small-world model in which edges are randomly added between pairs of selected vertices so that no edges are removed from the loop lattice \cite{b14}. The NW model enables networks to exhibit small-world characteristics by adding a small number of shortcuts into networks.

The small-world characteristics of wireless networks was studied in \cite{b15,b16,b17,b18,b19,b20}. In \cite{b15}, Zheng proposed a model to capture small-world characteristics of wireless networks. This model enables the flexible data rate in the proposed network model with social relationships which make nodes more likely to communicate with nearby nodes. In \cite{b16}\cite{b17}, the concept of small-world network was extended to a two-layer heterogeneous network in which shortcuts are added between data aggregators. The proposed heterogeneous network architecture enabled machines to communicate through shortcuts rather than the long concatenation of multi-hop transmissions to reduce the transmission delay and improve the throughput of the wireless network. A data dissemination method based on community perception was proposed in \cite{b18} by exploring the distinct small-world characteristics. It can reduce the delay and increase the transmission data rate. The correlation model of Base Station (BS) spatial traffic characteristics was established based on the large-scale measurement data set and numerical results showed that it exhibits the characteristics of fractal and small-world \cite{b19}. A small-world model of delay optimization was proposed for the BS caching network, which reduces the average delay by decreasing the average hop count \cite{b20}. The above studies explore small-world characteristics in designing wireless networks. But there have been fewer works in designing IoT routing schemes utilizing small-world characteristics in the open literatures.

There exist many studies investigating the delay of IoT networks \cite{b21,b22,b23,b24,b25,b26}. A routing scheme which aims to reduce the energy consumption as well as end-to-end delay of Industrial IoT systems was proposed in \cite{b21}. A new event-aware backpressure scheduling scheme was designed to enhance the real-time performance of IoT networks\cite{b22}. A differentiated service based data routing scheme was proposed in \cite{b23}, which intelligently selects the optimal routing according to the delay requirement of traffic. An unequal clustering algorithm was proposed to solve the problem of unbalanced energy consumption and delay in wireless sensor networks (WSNs) \cite{b24}. The delay in large scale networks have been analyzed by combining the stochastic geometry and queueing theory \cite{b25,b26}. Nevertheless, the above studies mainly focused on improving the delay and do not analyze the reliability of IoT networks. The reliability of IoT networks was studied in \cite{b6},\cite{b27,b28,b29}. An emergency response IoT routing protocol based on global information decision was proposed to ensure the reliability of data transmission and the efficiency of emergency response in IoT networks \cite{b6}. However, the proposed scheme makes decisions based on global information, leading to heavy computation cost. A joint clustering and routing protocol was proposed to achieve reliable data collection in large-scale WSNs and the relationship between clustering and routing was analyzed \cite{b27}. A new zone-based efficient routing protocol was proposed in \cite{b28} to realize efficient data transmission without affecting reliability. A heuristic ant colony algorithm combining local exploration was proposed to meet the requirement of a specified minimum reliability level at the lowest deployment cost in WSNs \cite{b29}. However, these works only focus on accommodating the reliability of IoT networks, and do not consider the problem of large end-to-end delay caused by inherent structures. Nevertheless, the delay also depends on the reliability of networks and it is difficult to reduce delay and improve network reliability simultaneously in IoT networks.

To tackle this problem, this paper proposes a new IoT routing mechanism considering small-world characteristics based on Cayley graphs \cite{b30,b31}. In terms of small-world characteristics of IoT networks, some shortcuts should be added into IoT networks to reduce the total delay. However, how to add shortcuts into IoT networks is a challenge. The shortest routing with uniform loads between nodes with shortcuts should be considered to achieve the low latency and reliability of IoT networks. Bringing Cayley graphs generated by transpositions into network modeling allows the shortest routing with uniform loads between nodes \cite{b31}. Shortcuts are added into IoT networks based on Cayley graphs which enables the nodes with shortcuts to communicate with each other through the shortest routing with uniform loads. In view of small-world characteristics, we analyze performance of the proposed IoT routing mechanism.

The main contributions of this paper are summarized as follows:

1) Based on the Cayley graph theory, a new IoT routing mechanism is proposed. ASL and ACC of IoT networks are derived to quantitatively analyze the performance of the small-world characteristics.

2) Based on ASL and ACC of IoT networks, the reliability and delay are investigated for Small World IoT based on Cayley grapH (SWITCH).

3) Compared with Nearest Neighboring Routing (NNR), the proposed SWITCH has lower delay and higher reliability than that of NNR. Specifically, the maximum delay of SWITCH is reduced by 50.6\%.

This paper is organized as follows: Preliminaries about Cayley graphs are introduced in section II. The system model of the IoT network based on Cayley graph is described in Section III. The ASL and ACC of IoT networks based on Cayley graph are derived in Section IV. Moreover, the small-world characteristics of IoT networks are validated by simulations. The reliability and delay of SWITCH are derived in Section V. Simulation results are given in Section VI. Finally, conclusions are drawn in Section VII.


\section{PRELIMINARIES}
In this section, the basics of the group and the Cayley graph are introduced. A group is an algebraic structure consisting of a set of elements that meet certain combinatorial rules. Combinatorial rules (often called multiplication) can be the expansion of ordinary multiplication, matrix multiplication or the interchange of two elements. In the permutation group $Q$, the elements are different permutations of $n$ integers $\{ 1,2,3 , \cdots , n \}$. There are $n!$ permutations in the permutation group $Q$. The Cayley graph $cay(Q)$ is generated by permutations on $\{ 1,2,3 , \cdots , n \}$, where the nodes in $cay(Q)$ are elements in $Q$ and edges connecting two nodes represent the interchange operations between two elements. The star graph is one kind of Cayley graphs. The nodes in the star graph $S _ { n }$ are in the interchange form of $\left( p _ { 1 } , p _ { s } \right)$ denoting that the symbol $p _ { s }$ is replaced by the symbol $p _ { 1 }$, where the symbol $p _ { 1 }$ represents the first number of the element and $p _ { s }$ represents the $s$-th number of the element \cite{b31}.

$S_{n, k}$ is a general expression of the star graph, where $n$ and $k$ $(1 \leq k \leq n-1)$ are parameters determining the size of the star graph. The number of nodes in the star graph $S_{n, k}$ is $N=n ! /(n-k) !$ and the node degree is $n-1$. A star graph $S_{n, k}$  can be regarded as a combination of $n$ mutually disjoint star subgraphs $S_{n-1, k-1}$. A node $V$ in the star graph $S_{n, k}$ can be expressed as $p_{1} p_{2} \cdots p_{s} \cdots p_{k}$ and the node set of the star graph $S_{n, k}$ is $\left\{p_{1} p_{2} \cdots p_{s} \cdots p_{k} | p_{s} \in\{1,2, \ldots, n\}, p_{s} \neq p_{t} \text { for } s \neq t\right\}$. The adjacent nodes of the node $V=p_{1} p_{2} \cdots p_{s} \cdots p_{k}$ are defined according to two rules, to be explained shortly later, and are expressed as $V_{s}$ and $ V_{x}$ respectively. The formation rules of $V_{s}$ and $ V_{x}$ are shown below\cite{b32}: 1) $V_{s}=p_{s} p_{2} \cdots p_{s-1} p_{1} p_{s+1} \cdots p_{k}(2 \leq s<k)$ (swapping $p_{1}$ with $p_{s}$); 2) $V_{x}=p_{x} p_{2} \cdots p_{k}$, where $p_{x} \in\{k+1, k+2, \cdots,n\}$. For example, the node $V$ in the star graph $S_{7, 4}$ is expressed as 1234. According to the first formation rule, the adjacent nodes of the node $V$ are 2134, 3214, 4213. According to the second formation rule, the adjacent nodes of the node $V$  are 6234, 7234, 5234. Hence, adjacent nodes of the node $V$ in the star graph $S_{7, 4}$ are 2134, 3214, 4213, 6234, 7234, 5234.

\section{SYSTEM MODEL}

\subsection{Deployment of nodes}
Nodes of IoT networks can be classified into two classes: one is the super nodes (SNs) which support long-range transmissions, the other is the regular nodes (RNs) which only support short-range transmissions. SNs usually have two or more radios and stronger hardware capabilities than that of RNs. For example, some IoT devices with stronger transmission power can be considered as SNs and other IoT sensors with smaller transmission power can be considered as RNs in intelligent plants. Aimed at the characteristic of the star graph that allows the shortest routing with uniform load between nodes, the topology of SNs is configured as the star graph in this paper. Nodes of the star graph are SNs and the connected edges of the star graph are communication links between SNs. Every SN in the IoT network has a unique identity number denoted by the corresponding element in the node set of the star graph $S_{n, k}$.

\begin{figure}[!h]
\centering
  \includegraphics[width=0.5\textwidth]{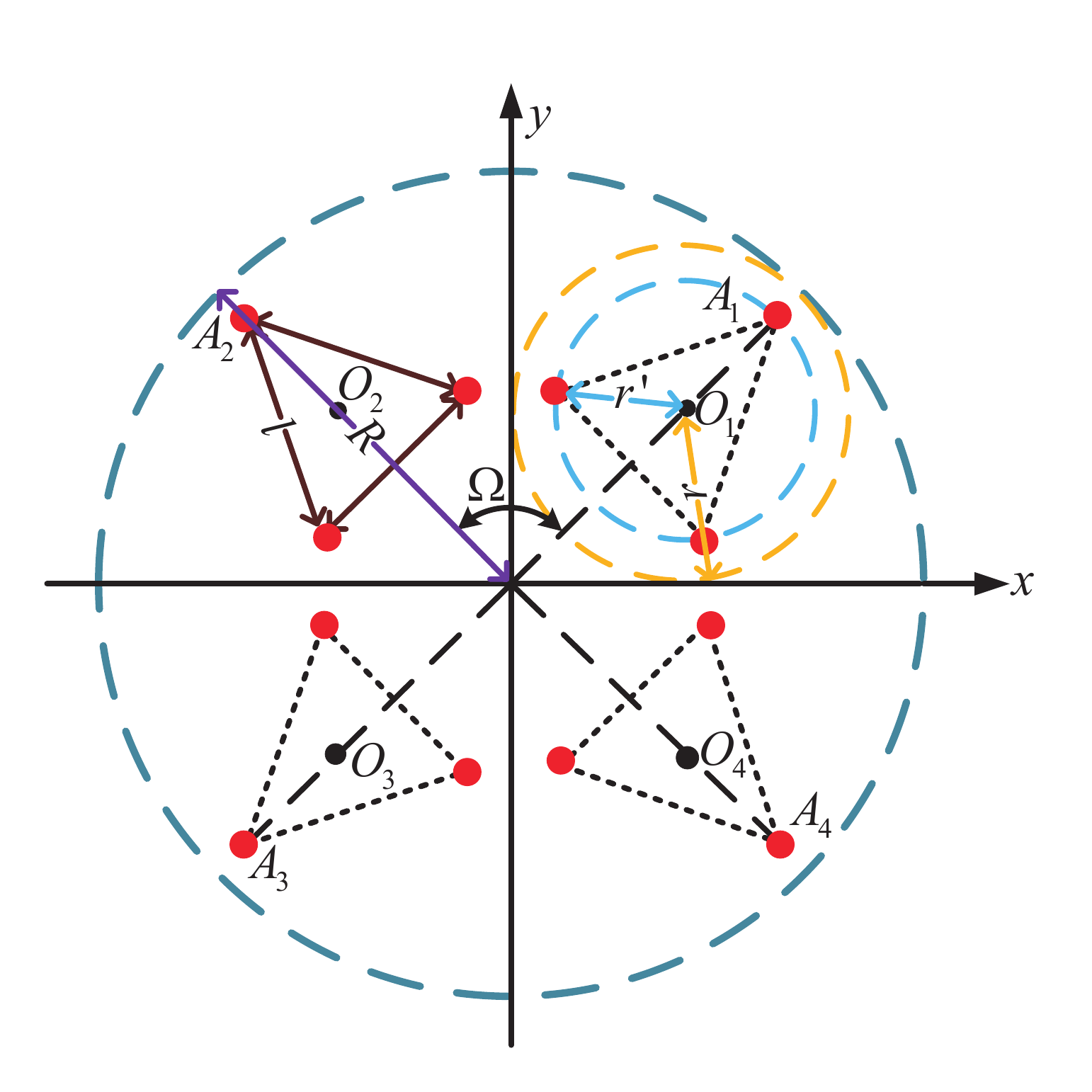}
\caption{Deployment of SNs (Taking the star graph $S_{4,2}$ as an example).}
\label{Fig1}       
\end{figure}

Consider the communication range, such as the deployment area of a smart factory. All SNs are deployed on a disk with the radius $R$, as shown in Fig. 1. According to the characteristics of the star graph that the star graph $S_{n, k}$ can be regarded as the combination of $n$ star subgraphs $S_{n-1, k-1}$, the circular communication area is divided into $n$ sectors with radian $\Omega=2 \pi / n$ and there are connected edges among sectors. The SNs of each sector are also deployed on an inner disk and the inner disk radius $r$ of each sector is given as
\begin{equation}
r=\frac{R \sin (\pi / n)}{1+\sin (\pi / n)}.
\end{equation}
In the two-dimensional Cartesian coordinate system shown in Fig. 1,  the deployment areas with the number of $n$ star subgraphs $S_{n-1, k-1}$ are the continuous and non-overlapping sectors centered at the origin with an angle $\Omega$. The coordinate of the inscribed disk center in the first sector can be given as $\mathrm{O}_{1}=(r / \tan (\pi / n), r)$. The inscribed disk center coordinates $\mathrm{O}_{m}(1<m \leq n)$ of other sectors can be obtained by counterclockwise rotating the coordinates of $\mathrm{O}_{1}$ by an angle of $\Omega \times(m-1)$. The radius of the inscribed disk is scaled by a proportion $\Delta $ in order to prevent the overlap of deployed nodes. The radius of the scaled disk is $r^{\prime}=\Delta \times r$ and the center coordinate of the scaled disk is the same as the center coordinate of the inscribed disk. SNs are deployed uniformly on the scaled disk with the center $\mathrm{O}_{m}$  and a radius $r^{\prime}$. For the entire deployment area, SNs are not distributed uniformly. The distance between adjacent nodes in each sector is $l=2 r^{\prime} \sin (\pi / n)$. The node $A_{1}$ in the first sector is placed on the extension line connecting the origin and the point $\mathrm{O}_{1}$. The convenience of this process is that $A_{1}$ can rotate the same angle as $\mathrm{O}_{1}$ to obtain $A_{m}$ and $\mathrm{O}_{m}$ in other sectors.
\begin{figure}[!h]
\centering
  \includegraphics[width=0.5\textwidth]{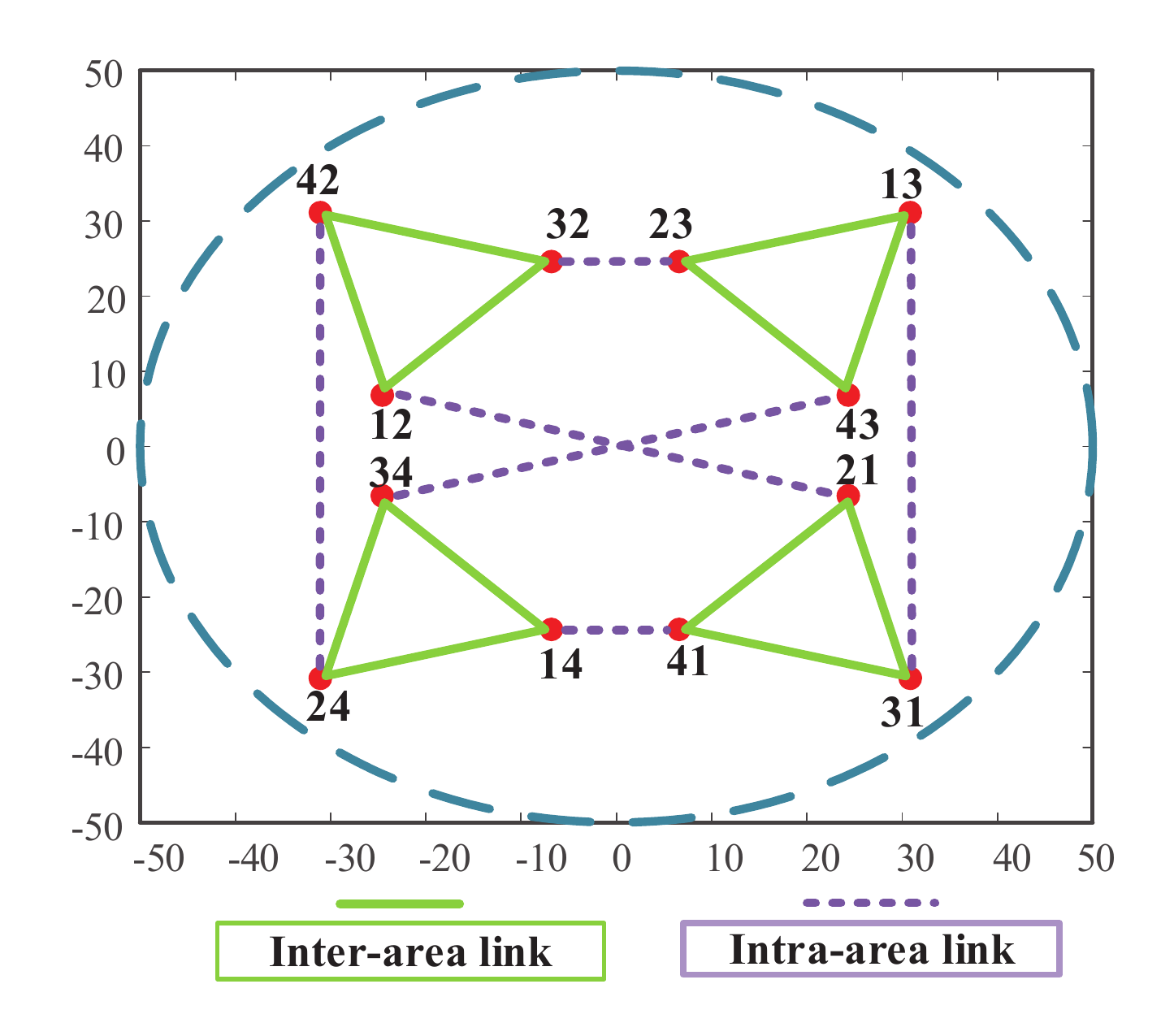}
\caption{ Communication links between SNs (Taking the star graph $S_{4,2}$ as an example).}
\label{Fig2}       
\end{figure}

Fig. 1 shows the deployment of SNs, which takes the star graph $S_{4, 2}$ as an example. According to the star graph $S_{4, 2}$,  the entire circular communication area is divided into four sub-areas, each of which is a sector with a radian of $\pi / 2$. In each sub-area, there are three SNs which are uniformly distributed on the disk with $\mathrm{O}_{m}$  as the center and $r^{\prime}$ as the radius. Therefore, the locations of SNs can be obtained if the center and radius of the circular region as well as the structure of the corresponding star graph are determined. Meanwhile, each SN is assigned an identity number, and the identity numbers of SNs within the same sub-area have the same value of the last number. Finally, the topological relationship between SNs is established based on the connection rules of the star graph. Fig. 2 shows the connection situation after SNs are deployed based on the star graph $S_{4, 2}$ where the identity numbers are randomly assigned to the SNs in each sub-area. Considering that the star graph allows the shortest routing with uniform load between nodes \cite{b31}, SNs can communicate with each other through the shortest routing with uniform load in Fig. 2. Consider that only a few SNs need to be deployed in the communication area based on the proposed topology, smart factories can deploy SNs in their plants to help IoT sensors transmit in a long distance.

\subsection{IoT Network Framework}
The system model of IoT network is shown in Fig. 3. SNs are selected based on the formation rules of the star graph $S_{n, k}$. SNs are fixed in a disk with the radius $R$. Hence, the star graph only includes SNs, each with a unique identity. Moreover, in this paper SNs are configured with the long-range transmission capability and RNs are configured with the short-range transmission capability. In Fig. 3, the connected edges represent communication links between SNs. RNs are assumed to be governed by a homogeneous Poisson point process with parameter $\lambda_{\rm R N}$ \cite{b9}. When the source regular node (SRN) communicates with the destination regular node (DRN), the SRN first needs to determine whether to adopt SWITCH or the traditional NNR. If the SRN and DRN are very close to each other and the number of hops with NNR is smaller than that obtained using SWITCH, the SRN transmits to the DRN directly. Otherwise, the SRN first selects the nearest SN named source super node (SSN) for data transmission. The SSN transmits data to the destination super node (DSN) which is closest to the DRN. Finally, the DSN transmits data to the DRN. Based on the star graph in Fig. 3, the shortest routing with uniform load among SNs is established for data transmission.
\begin{figure}[!h]
\centering
  \includegraphics[width=0.5\textwidth]{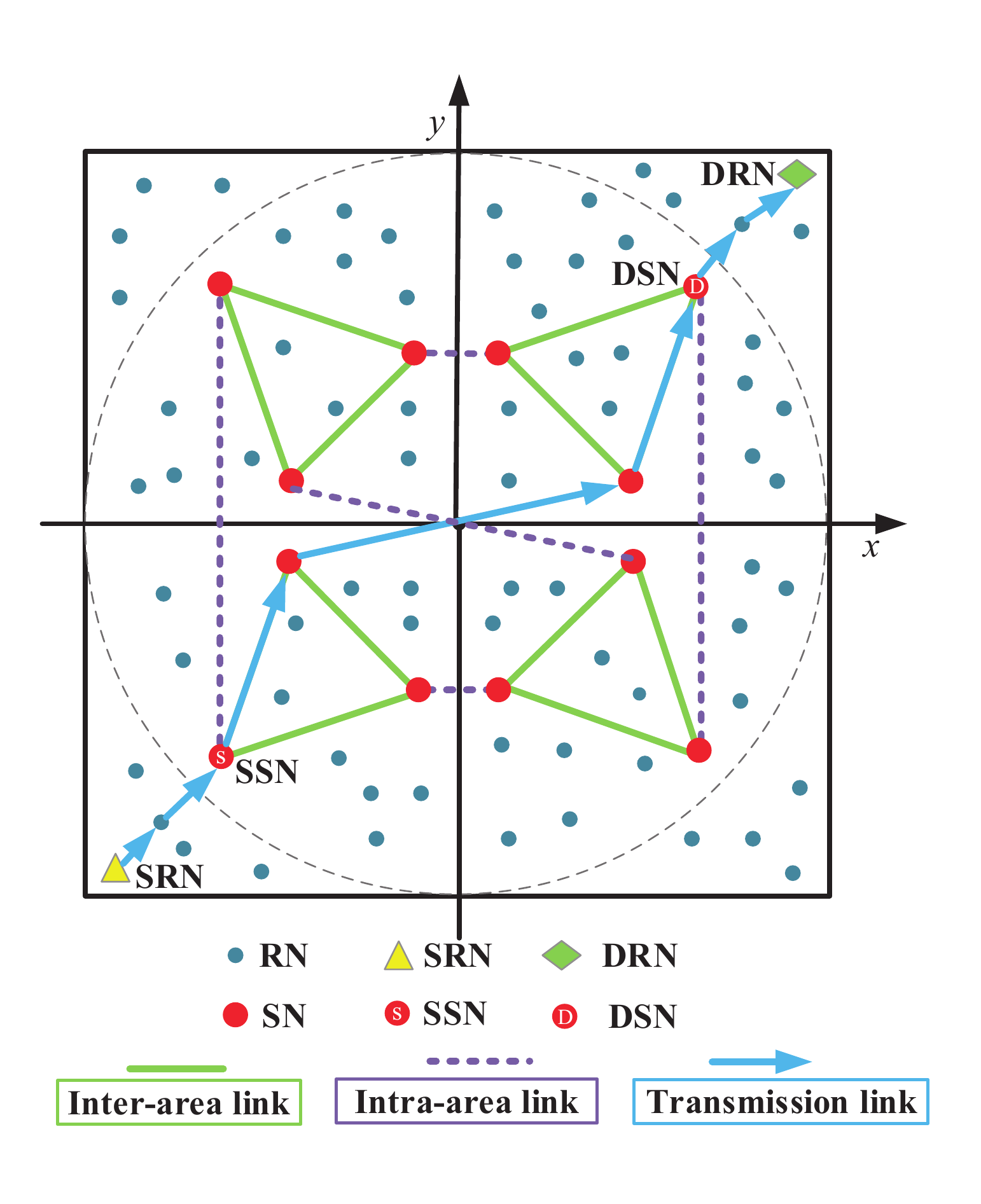}
\caption{IoT communication network framework based on star graph $S_{n, k}$.}
\label{Fig3}       
\end{figure}
\subsection{Routing Strategy}
Based on the system model in Fig. 3, the data of SRN is first transmitted to the nearest SSN through multi-hop transmissions. One of typical multi-hop schemes is the NNR scheme \cite{b29}. Considering only the short-range communication capability in RNs, the NNR scheme is adopted for transmissions between SRN and SSN. Furthermore, the routing from SSN to DSN can be established by the star graph $S_{n, k}$ considering the long-range communication capability of SNs. Based on the node symmetry of the star graph $S_{n, k}$, an identity number of a SN $E_{\rm t}$ in the Fig. 3 can be mapped to the new identity number of the SN which is denoted as $E_{\rm k}$ by the renaming function \cite{b31}. Considering routings between the SSN and DSN, the identity number of DSN $E_{\rm DSN}$  is mapped to a new identity number of SN $E_{\rm k}^{\rm DSN}=1,2, \ldots, k$ by the renaming function $M(\bullet)$, i.e., $M\left(E_{\rm DSN}\right)=E_{\rm k}^{\rm DSN}$ and the identity number of SSN is mapped to a new identity number of renamed SSN denoted as $M\left(E_{\rm SSN}\right)=E_{\rm k}^{\rm SSN}$. Routings between the SSN and DSN in the original graph are isomorphic to routings between the SN with the identity number $E_{\rm k}^{\rm SSN}$ and another SN with the identity number $E_{\rm k}^{\rm DSN}$ in the renamed graph. Based on the shortest routing scheme of star graphs \cite{b32}, the minimum hop number between the SSN and DSN can be obtained. The DSN transmits data to the DRN based on the NNR scheme after the data arrives at the DSN. In the end, the data is transmitted from the SRN to the DRN.
\subsection{Channel Model}
In this paper, a simple one-packet one-transmission (1P1T) protocol is configured for IoT networks \cite{b34}.  All nodes are required to synchronize and adopt the time division multiple access protocol. That is, only one node i.e., RN or SN can transmit in each time slot over the source-relay-destination link. Moreover, the channels of SNs are orthogonal to each other. Hence, a node can only be interfered by other RNs which are not located along the source-relay-destination link\cite{a2}. The data transmission from the node ${u}$ to the node  ${v}$ is successful if the signal-to-interference-plus-noise ratio (SINR) of receiver node is higher than a specified threshold. The SINR of the receiver node is expressed as
\begin{equation}
\gamma_{u v}=\frac{P_{u} d_{u v}^{-\alpha}\left|h_{u v}\right|^{2}}{P_{\rm I} \sum_{a \in \Upsilon_{v}} I_{a v}+\sigma_{z}^{2}},
\end{equation}
where $P_{u}$ is the transmit power of node $u$, $d_{u v}$  is the distance between nodes $u$ and $v$, $\alpha$ is the path loss exponent, $I_{a v}$ is the interference received by node $v$, $\sigma_{z}^{2}$ is the noise power, $P_{\rm I}$  is the transmission power from interference nodes which are assumed to be the same for all interference nodes, $h$ is the small-scale fading coefficient with zero mean complex Gaussian distribution of unit variance and its amplitude follows a Rayleigh distribution, $\Upsilon_{v}$ is a set of inter-route interference nodes from RNs.

\section{SMALL-WORLD CHARACTERISTICS OF IOT BASED ON CAYLEY GRAPHS}
The IoT networks with small-world characteristics have small ASL and high ACC \cite{b12}. A small ASL means that a few hops are configured to transmit data between two nodes, which reduces delay and energy consumption in IoT networks. The high ACC means that the data can be spreaded widely across IoT networks \cite{b11}. In this section, the small-world characteristics of IoT networks based on Cayley graphs are investigated by being compared with the IoT networks without the added shortcuts. Moreover, the ASL and ACC of SWITCH are derived for performance analysis.

\subsection{Average Shortest-path Length}
The ASL is the average number of hops required for data transmission from the source node to the destination node. Based on Fig. 3, the NNR scheme and the star graph $S_{n, k}$ are adopted to form a new routing for the proposed small-world IoT based on Cayley graphs. The NNR scheme aims to guarantee the link quality of every hop by minimizing the hop distance \cite{b33}. The RN always chooses the nearest RN in the sector with the angle $\Phi$ as the next hop along the direction to the destination. In this paper, the RNs are assumed to be governed by a homogeneous Poisson point process with parameter $\lambda_{\rm RN}$ \cite{a3}. The distance between the RN and its nearest RN in the sector with the angle $\Phi$ is denoted as $r_{\rm RN}$.  The probability distribution function of $r_{\rm RN}$ is expressed as \cite{b33}

\begin{equation}
\operatorname{Pr}\left\{\mathrm{r}_{\rm R N} \leq x\right\}=1-e^{-\frac{\lambda_{\rm R N} \Phi x^{2}}{2}}.
\end{equation}
The average value of $r_{\rm RN}$ is $\mathbb{E}\left(r_{\rm R N}\right)=\sqrt{\frac{\pi}{2 \lambda_{\rm R N} \Phi}}$. The distance of each hop projected to the straight line between the SRN and DRN is denoted as $X$, and the average value of $X$  is expressed as

\begin{equation}
\begin{aligned}\mathbb {E}(\mathrm{X}) &=\mathbb {E}\left(r_{\rm RN}\right) \eta(\Phi) \\ &=\mathbb {E} \left(r_{\rm RN}\right) \frac{2}{\Phi} \sin \left(\frac{\Phi}{2}\right) \end{aligned},
\end{equation}
where $\eta(\Phi)=\mathbb {E}[\cos \phi]=\frac{2}{\Phi} \int_{0}^{\frac{\Phi}{2}} \cos \phi d \phi=\frac{2}{\Phi} \sin \frac{\Phi}{2}$ is the path efficiency, representing the ratio of Euclidean distance to actual propagation distance \cite{b35}.  The parameter $\phi(|\phi| \leq \Phi / 2)$ is the emission angle of the source-destination link. $L_{1}$ and $W_{1}$ are the distance and hop number between the SRN and SSN, respectively. $L_{2}$ and $W_{2}$ are the distance and hop number between the DRN and DSN, respectively. $W_{1}$ and $W_{2}$ are expressed as

\begin{equation}
W_{1} = \frac{L_{1}}{\mathbb {E}\left(r_{\rm RN}\right) \frac{2}{\Phi} \sin \left(\frac{\Phi}{2}\right)},
\end{equation}

\begin{equation}
W_{2} = \frac{L_{2}}{\mathbb {E}\left(r_{\rm RN}\right) \frac{2}{\Phi} \sin \left(\frac{\Phi}{2}\right)}.
\end{equation}

The minimum number of hops from SSN to DSN can be obtained based on the shortest routing scheme of the star graph $S_{n, k}$ \cite{b32}. The star graph $S_{n, k}$ has the shortest routing with a uniform load, which should be analyzed to explain why the star graph $S_{n, k}$ is used to establish routings among SNs. The vertex forwarding factor is used to measure the maximum paths going through the node which is also called the load of the node when the shortest routing is adopted. In the Cayley graph $cay$, there exists a shortest routing in which the load of each node $v$ known as the vertex forwarding factor is equal to $\zeta(\mathrm{Cay})=\sum_{v, u \in V, v \neq u} w(u, v)-(n-1)$, where $w(u, v)$ refers to the hop number from the node $u$ to $v$ \cite{b31}. $\zeta($ Cay $)$ is a constant for any nodes in the Cayley graph when $n$ is determined, which explains that the Cayley graph has the shortest routing with a uniform load between nodes. Considering SNs selected by Cayley graph in SWITCH, there exists a shortest routing with the uniform load among SNs. The maximum value of the minimum hop number between two nodes in the star graph $S_{n, k}$ is expressed as \cite{b32}

\begin{equation}
D(n, k)=\left\{\begin{array}{c}{2 k-1, k \leq\lfloor n / 2\rfloor} \\ {\lfloor(n-1) / 2\rfloor+ k, k \geq\lfloor n / 2\rfloor+ 1}\end{array}\right.
\end{equation}
The number of transmission hops from SSN to DSN can be denoted as $D(n, k)$. Then, the ASL of SWITCH is expressed as
\begin{equation}
W_{\rm SWITCH}=W_{1}+W_{2}+D(n, k).
\end{equation}
Since the distance between the SRN and DRN is 2$\sqrt{2} R$, the ASL of NNR is expressed as
\begin{equation}
W_{\rm NNR} = \frac{2 \sqrt{2} R}{\mathbb {E}\left(r_{\rm RN}\right) \frac{2}{\Phi} \sin \left(\frac{\Phi}{2}\right)}.
\end{equation}

\subsection{Average Clustering Coefficient}
The clustering coefficient indicates the clustering degree of nodes in IoT networks. Supposing that the node $v$ in the IoT network has $b$ neighbor nodes which can communicate with node $v$. Many communication links exist among neighbor nodes of node $v$ when neighbor nodes of node $v$ can communicate with each other. The number of communication links existing among neighbor nodes of the node $v$ is denoted as $\delta$. The clustering coefficient of the node $v$  is expressed as \cite{b20}
\begin{equation}
C_{v}=\frac{2 \delta}{b(b-1)}.
\end{equation}
Then, the ACC of IoT networks can be expressed as
\begin{equation}
C_{\rm av}=\frac{1}{M} \sum_{v=1}^{M} C_{v},
\end{equation}
where $M$ is the total number of nodes in the IoT network.

Corollary 1$:$ In NNR, the ACC can be expressed as
\begin{equation}
\begin{array}{l}
{C_{\rm NNR}} = \int\limits_0^{\mathbb{E}({r_{\rm RN}})} {\left[ {\frac{2}{\pi }\arccos \left( {\frac{{{x_{\rm d}}}}{{2\mathbb{E}({r_{\rm RN}})}}} \right)} \right.} \\
{\kern 1pt} {\kern 1pt} {\kern 1pt} {\kern 1pt} {\kern 1pt} {\kern 1pt} {\kern 1pt} {\kern 1pt} {\kern 1pt} {\kern 1pt} {\kern 1pt} {\kern 1pt} {\kern 1pt} {\kern 1pt} {\kern 1pt} {\kern 1pt} {\kern 1pt} {\kern 1pt} {\kern 1pt} {\kern 1pt} {\kern 1pt} {\kern 1pt} {\kern 1pt} {\kern 1pt} {\kern 1pt} {\kern 1pt} {\kern 1pt} {\kern 1pt} {\kern 1pt} {\kern 1pt} \left. {{\kern 1pt}  - \frac{{{x_{\rm d}}\sqrt {4{{[\mathbb{E}({r_{\rm RN}})]}^2} - {x_d}^2} }}{{2\pi {{[\mathbb{E}({r_{\rm RN}})]}^2}}}} \right]\frac{{2{x_{\rm d}}}}{{{{[\mathbb{E}({r_{\rm RN}})]}^2}}}d{x_{\rm d}}
\end{array},
\end{equation}
where $x_{\rm d}$ is the distance from the RN $v_{\rm R}$ to neighbor RNs.

Proof$:$ See Appendix A

Theorem 1: If the number of SNs deployed in the IoT network is $N$ , then the ACC of SWITCH can be expressed as
\begin{equation}
\begin{array}{l}{C_{\rm SWITCH}} \\ {=\frac{M_{\rm R} C_{\rm NNR}}{M_{\rm R}+N}+\frac{2 N\left(\delta_{\rm R}+e_{\rm S}\right)}{\left(b_{\rm R}+n-1\right)\left(b_{\rm R}+n-2\right)\left(M_{\rm R}+N\right)}}\end{array},
\end{equation}
where $M_{\rm R}$ is the number of RNs in the IoT network, $e_{\rm S}\left(0 \leq e_{\rm S} \leq n-2\right)$ is the number of communication links existing among neighbor SNs of the SN $v_{\rm S}$, $b_{\rm R}$ is the number of neighbor RNs of the RN $v_{\rm R}$, $\delta_{\rm R}$ is the number of communication links existing among neighbor RNs of the RN $v_{\rm R}$.

Proof$:$ See Appendix B
\begin{figure}[!h]
\centering
  \includegraphics[width=0.5\textwidth]{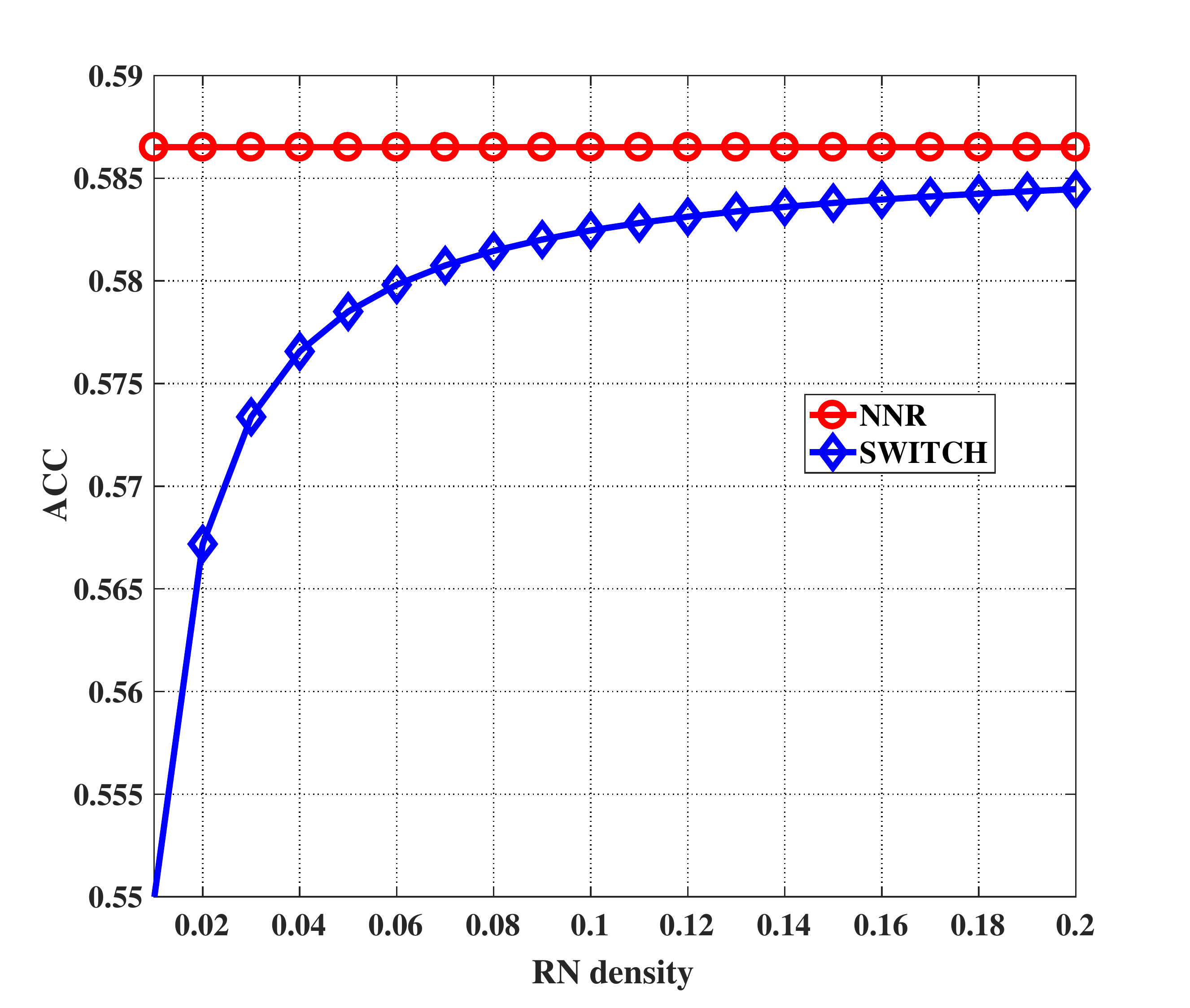}
\caption{RN density as function of ACC based on SWITCH and NNR.}
\label{Fig41}       
\end{figure}

\begin{figure}[!h]
\centering
  \includegraphics[width=0.5\textwidth]{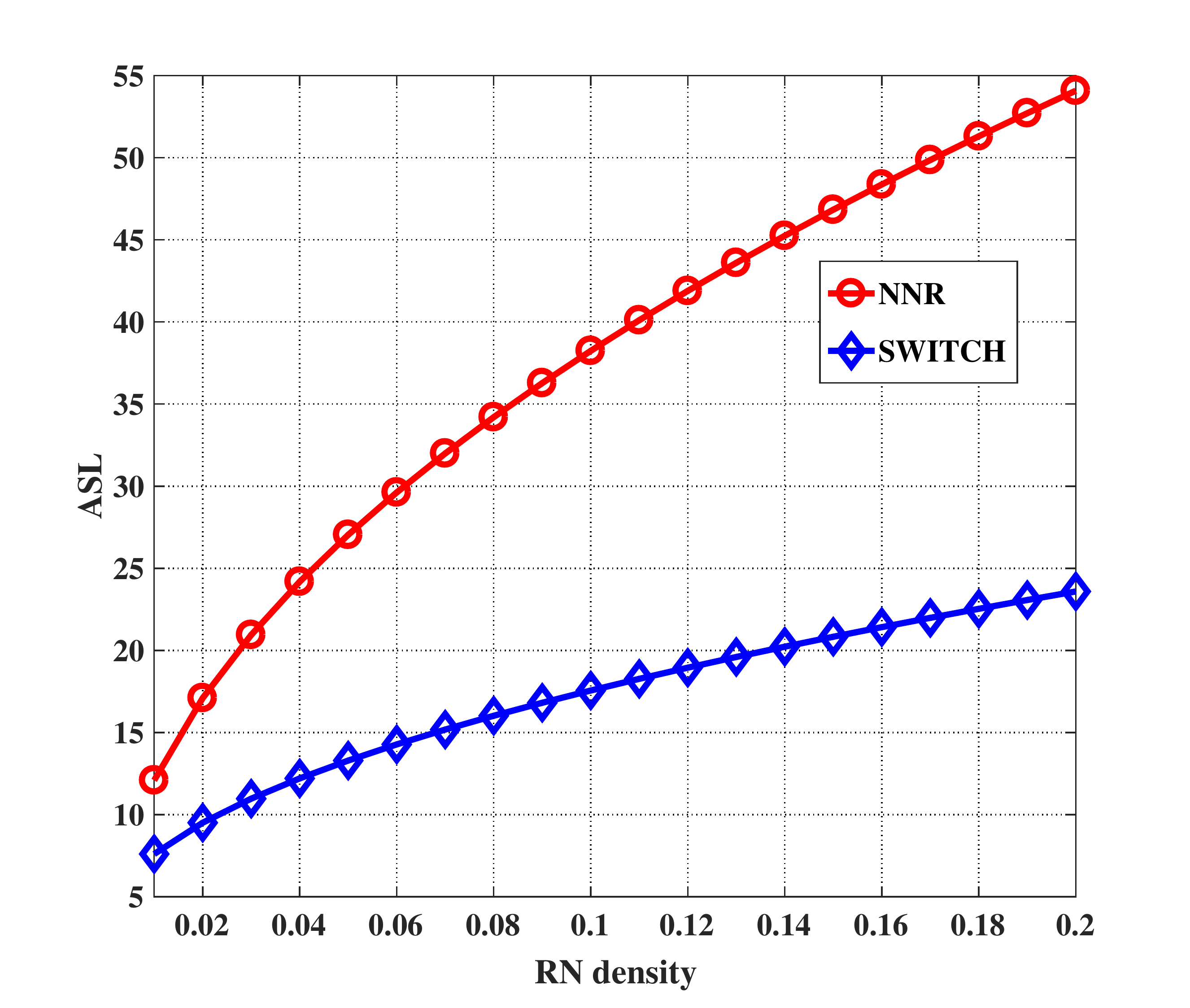}
\caption{RN density as function of ASL based on SWITCH and NNR.}
\label{Fig5}       
\end{figure}
Fig. 4 shows the RN density as function of the ACC based on SWITCH and NNR. The ACC of SWITCH increases with the RN density $\lambda_{\rm RN}$. The ACC of NNR is independent on the RN density $\lambda_{\rm RN}$. When the RN density is fixed, the ACC of SWITCH is less than the ACC of NNR. When the RN density $\lambda_{\rm RN}$ increases, the ACC of SWITCH gradually approaches to the ACC of NNR.

Fig. 5 shows the RN density as function of the ASL based on SWITCH and NNR. The ASL increases with the RN density $\lambda_{\rm RN}$. The ASL of NNR is obviously higher than the ASL of SWITCH. The gap between the ASL of NNR and that of SWITCH grows with the RN density $\lambda_{\rm RN}$.
Based on the results in Fig. 4 and Fig. 5, the ACC of SWITCH is approximately equal to that of NNR and the ASL of SWITCH is obviously lower than that of NNR when the RN density $\lambda_{\rm RN}$ grows. Considering the distinguishing features of small-world networks, SWITCH exhibits small-world characteristics.

\section{MODELING OF RELIABILITY AND DELAY}
In this section, the reliability and delay models of SWITCH are derived based on the ASL and ACC of IoT networks.
\subsection{Reliability Model}
\subsubsection{The success probability}
The success probability of data transmission is the probability that the receiving node successfully decodes the data from the transmitting node. The data can be successfully decoded if the SINR of the receiving node is larger than or equal to a given threshold $\beta$. The success probability of data transmission at the receiving node is expressed as
\begin{equation}
\rho\left(d_{u v}\right)=\operatorname{Pr}\left(\gamma_{u v} \geq \beta\right).
\end{equation}
The interference received at the receiving node $j$ is expressed as
\begin{equation}
I_{v}=\sum_{a \in \Upsilon_{v}} d_{av}^{-\alpha}\left|h_{av}\right|^{2},
\end{equation}
where $\Upsilon_{v}$ is a set of inter-route interference nodes which are assumed to be governed by a uniform Poisson point process with parameter $\lambda_{\rm I}$. The success probability of data transmission at the receiving node $v$ is expressed as \cite{b34}
\begin{equation}
\begin{aligned}
&\rho\left(d_{u v}, \Upsilon_{v}, \mathrm{P}_{u}\right)=\operatorname{Pr}\left(\gamma_{uv} \geq \beta\right)
\\&=\operatorname{Pr}\left\{\left|\mathrm{h}_{u v}\right|^{2}<\frac{\beta d_{u v}^{\alpha}\left(\sum_{a \in \Upsilon_{ v}} P_{\rm I} I_{ av}+\sigma_{z}^{2}\right)}{P_{u}}\right\}
\\&=\operatorname{Pr}\left\{\left|\mathrm{h}_{u v}\right|^{2}<\frac{\beta d_{uv}^{\alpha} \sigma_{z}^{2}}{P_{u}}+\frac{\sum_{a \in \Upsilon_{v}} P_{\rm I} I_{av} \beta d_{ uv}^{\alpha}}{P_{ u}}\right\}
\\&=\exp \left(-\frac{\beta d_{uv}^{\alpha} \sigma_{z}^{2}}{P_{u}}\right) \times \mathbb {E}_{Iv}\left[\prod_{a \in \Upsilon_{v}} \exp \left(-\frac{P_{\rm I} I_{av} \beta d_{uv}^{\alpha}}{P_{u}}\right)\right]
\\&=\exp \left(-\frac{\beta d_{uv}^{\alpha} \sigma_{z}^{2}}{P_{u}}-\lambda_{\rm I} q_{\alpha}\left(\frac{P_{\rm I}}{P_{u}} \beta\right)^{\frac{2}{\alpha}} d_{uv}^{2}\right)
\end{aligned},
\end{equation}
where $q_{\alpha}=\pi \Gamma(1+(2 / \alpha)) \Gamma(1-2 / \alpha)$.

In SWITCH, the channels of SNs are orthogonal to each other. The interference received at the receiving nodes is mainly the inter-route interference from the RNs which are not located at the source-relay-destination link. The transmitting power of RNs is $P_{\rm RN}$. The transmitting power of SNs is configured as $P_{\rm SN}=\varepsilon P_{\rm R N}(\varepsilon>0)$. The distance $L_{\rm l o n g}$ between SNs in different sub-areas is within the region $0<L_{\rm long} \leq 2 R(1+\Delta \sin (\pi / n)) /(1+\sin (\pi / n))$. The distance between SNs in different sub-areas has little effect on the success probability of data transmission, because the noise power over channels is small and the interference is ignored between SNs. Hence, $L_{\rm long}$ is approximately to the maximum value and is expressed as $L _ {\rm long } = 2 R ( 1 + \Delta \sin ( \pi / n ) ) / ( 1 + \sin ( \pi / n ) )$ in this paper. The number of transmission hops within the sub-area is denoted as $c _ { 1 }$ and the number of transmission hops between sub-areas is denoted as $c _ { 2 }$. In both cases of transmissions between RNs and transmissions from RN to SN, the receiving nodes suffer from inter-route interference. In both cases of transmissions between SNs and transmissions from SN to RN, the inter-route interference can be ignored due to orthogonality between SNs \cite{a4}. Thus, the total success probability of SWITCH is given by
\begin{equation}
\begin{aligned}
 \rho_{\text {SWTTCH}} & =\left[\rho\left(\mathbb {E}\left(r_{\rm RN}\right), \Upsilon_{v}, P_{\rm RN}\right)\right]^{W_{1}+W_{2}-1} \\ & \times \rho\left(\mathbb {E}\left(r_{\rm RN}\right), \Theta, P_{\rm SN}\right) \times\left[\rho\left(l, \Theta, P_{\rm SN}\right)\right]^{c_{1}}\\&\times\left[\rho\left(L_{\text {long}}, \Theta, P_{\rm SN}\right)\right]^{c_{2}}
\end{aligned},
\end{equation}
where $\Theta$ is an empty set. In NNR, $W _ {\rm N N R }$ is the ASL between the SRN and DRN. The total success probability of NNR can be expressed as
\begin{equation}
\rho_{\rm NNR}=\rho\left(\mathbb {E}\left(r_{\rm RN}\right), \Upsilon_{v}, P_{\rm RN}\right)^{W_{\rm NNR}}.
\end{equation}
\subsubsection{Reliability factor}
In this paper, the reliability factor of IoT networks is configured as the upper bound of the probability that each hop succeeds in decoding data when at most $G$ retransmissions are allowed for each data packet. The transmitting node is assumed to transmit data with the probability $\ell$ which is independent of other nodes. When the ACC of node $v$  is large, the nodes communicating with the node $v$ have a large probability to communicate with each other based on the definition of ACC. Moreover, the node $v$ with a higher ACC has a larger probability $\ell$ to transmit data. Hence, the probability $\ell$ is proportional to the ACC value of nodes and can be expressed as $\ell = \varpi C _ {\rm  a v } ( \varpi > 0 )$. The upper bound of the probability that the receiving node successfully decodes data with at most $G$ retransmissions is given by \cite{b36}
\begin{equation}
P _ {\rm s , \max } = 1 - ( \ell q + 1 - \ell ) ^ { G + 1 },
\end{equation}
where $q = 1 - \rho \left( d _ {u v } , \Upsilon _ { v } , \mathrm { P } _ { u } \right)$.

In IoT networks, the number of faulty nodes is denoted as $f$ which increases over time with a constant failure rate $a$ \cite{b37}. Hence, $f$ is expressed as \cite{b37}
 \begin{equation}
f = M \left( 1 - e ^ { - a \tau T _ {\rm s } } \right),
\end{equation}
where $T _ { \rm s }$ is the number of time intervals. Each time interval is assumed to include $\tau$ consecutive time slots. The non-failure probability of nodes is derived by
\begin{equation}
\Gamma = 1 - \frac { f } { M } = e ^ { - a \tau T _ {\rm s } }.
\end{equation}
For each hop, the reliability factor is expressed as
\begin{equation}
\mathrm { Re } = P _ { \rm s , \max } \Gamma.
\end{equation}

In SWITCH, the topology of SNs has the shortest routing with a uniform load. In this case, SNs are assumed to have the same non-failure probability of $\Gamma _ {\rm S N } = e ^ { - a _ { 1 } \tau T _ {\rm s } }$. RNs are assumed to be governed by a homogeneous Poisson point process in this paper. The non-failure probability of each RN is assumed to be equal to $\Gamma_{\rm R N}=e^{-a_{2} \tau T_{\rm s}}$. The reliability factor of SWITCH is
\begin{equation}
\begin{aligned} \mathrm{Re}_{\rm SWITCH} &=\left(P_{\rm s, \max }^{\rm R N-R N} \Gamma_{\rm R N}\right)^{W_{1}+W_{2}-1} \\ & \times\left(P_{\rm s, \max }^{\rm S N-S N, l}\right)^{c_{1}} \times\left(P_{\rm s, \max }^{\rm S N-S N, L_{\rm l o n g}}\right)^{c_{2}} \\ & \times P_{\rm s, \max }^{\rm S N-R N}\left(\Gamma_{\rm S N}\right)^{c_{1}+c_{2}+1} \end{aligned},
\end{equation}
where $P _ { \rm s , \max } ^ {\rm R N - R N }$,$P _ {\rm s , \max } ^ {\rm S N - S N , l }$,$P _ {\rm s , \max } ^ {\rm S N - S N , L _ {\rm l o n g } }$ and $P _ {\rm S , \max } ^ {\rm S N - R N }$ are the upper bounds of probabilities for receiving nodes to successfully decode data in four cases, i.e., transmissions between RNs, transmissions between SNs in inter-subarea, transmissions between SNs in intra-subarea and transmissions between the SN and RN. Similarly, the reliability factor of NNR can be expressed as
\begin{equation}
\mathrm{Re}_{\rm NNR}=\left(P_{\rm s, \max }^{\rm N N R} \Gamma_{\rm R N}\right)^{W_{\rm N N R}},
\end{equation}
where $P _ {\rm s , \max } ^ { \mathrm {\rm NNR } }$ is the upper bound of probability that the RN successfully decodes data with at most $G$ retransmissions.
\subsection{Delay Model}
The delay from the transmitting node to the next hop is denoted as $t _ { 1 }$. The delay of the acknowledgement (ACK) or negative acknowledgement (NACK) signals transmitted from the receiving node to the transmitting node is denoted as $t _ { 2 }$. Then, the delay of a retransmission is expressed as $t _ { 1 } + t _ { 2 }$, where $t _ { 1 }$ and $t _ { 2 }$ include the transmission delay and processing delay\cite{a5}. When the number of retransmissions from the node $u$ to the node $v$ is denoted as $C _ { u v }$, the average delay of one hop is expressed as
\begin{equation}
T _ {\rm t } \left( \mathrm { d } _ {u v } \right) = \mathbb {E} \left[ C _ {u v } \right] \mathrm { t } _ { 1 } + \left( \mathbb {E} \left[ C _ {u v } \right] - 1 \right) \mathrm { t } _ { 2 },
\end{equation}
where $\mathbb {E} \left[ \mathrm { C } _ {u v } \right]$ is the average value of $C _ {u v }$. In the ARQ mechanism, the number of retransmissions $C _ {u v }$ is a geometric random variable for each hop.  For the given distance $\mathrm{d}_{ u v}$, $\mathbb {E} \left[ C _ {u v} \right]$ can be expressed as
\begin{equation}
\mathbb {E} \left[ C _ {u v} \right] = \frac { 1 } { \rho \left( d _ {u v} \right) }.
\end{equation}
Substituting (26) into (25), the single hop delay in IoT networks is given by
\begin{equation}
T_{\rm t}\left(\mathrm{d}_{u v}\right)=\frac{t_{1}+t_{2}}{\rho\left(d_{u v}\right)}-\mathrm{t}_{2}.
\end{equation}
For a given multi-hop link of IoT networks, the average delay is
\begin{equation}
T_{\rm t}(\mathrm{d})=W\left(\frac{t_{1}+t_{2}}{\rho\left(d_{ u v}\right)}-t_{2}\right),
\end{equation}
where $W$ is the ASL.

In SWITCH, the delay from the SRN to the SSN is denoted as $T_{1}$, the delay from the SSN to the DSN is denoted as $T_{2}$ and the delay from the SRN to the SSN is denoted as $T_{3}$. $T_{1}$,$T_{2}$ and $T_{3}$ are given by

\begin{equation}
T_{1}=W_{1} \frac{t_{1}+t_{2}}{\rho\left(\mathbb {E}\left(r_{\rm R N}\right), \Upsilon_{v}, P_{\rm R N}\right)}-W_{1} t_{2},
\end{equation}

\begin{equation}
T_{2}=\frac{c_{1}\left(t_{1}+t_{2}\right)}{\rho\left(l, \Theta, P_{\rm S N}\right)}+\frac{c_{2}\left(t_{1}+t_{2}\right)}{\rho\left(L_{\rm long}, \Theta, P_{\rm SN}\right)}-\left(c_{1}+c_{2}\right) t_{2},
\end{equation}

\begin{equation}
\begin{aligned} T_{3}=& \frac{\left(W_{2}-1\right)\left(t_{1}+t_{2}\right)}{\rho\left(\mathbb {E}\left(r_{\rm R N}\right), \Upsilon_{v}, P_{\rm R N}\right)}+\frac{t_{1}+t_{2}}{\rho\left(\mathbb {E}\left(r_{\rm R N}\right), \Theta, P_{\rm S N}\right)} \\ &-W_{2} t_{2} \end{aligned}.
\end{equation}
The delay of SWITCH is

\begin{subequations}
\begin{equation}
\begin{aligned}
T_{\rm S W I T C H} &=T_{1}+T_{2}+T_{3} \\ &=\frac{\left(W_{1}+W_{2}-1\right)\left(t_{1}+t_{2}\right)}{\rho\left(\mathbb {E}\left(r_{\rm R N}\right), \Upsilon_{ v}, P_{\rm R N}\right)}+\frac{c_{1}\left(t_{1}+t_{2}\right)}{\rho\left(l, \Theta, P_{\rm S N}\right)} \\&+\frac{c_{2}\left(t_{1}+t_{2}\right)}{\rho\left(L_{\rm l o n g}, \Theta, P_{\rm S N}\right)}+\frac{t_{1}+t_{2}}{\rho\left(\mathbb {E}\left(r_{\rm R N}\right), \Theta, P_{\rm S N}\right)}\\&-W_{\rm S W I T C H} t_{2}
\end{aligned},
\end{equation}

\begin{equation}
W_{\rm S W I T C H}=W_{1}+W_{2}+c_{1}+c_{2}.
\end{equation}
\end{subequations}
The ASL of NNR is derived by (9), the delay of NNR can be derived by substituting (9) into (28) and then be expressed as
\begin{equation}
T_{\rm N N R}=W_{\rm N N R}\left(\frac{t_{1}+t_{2}}{\rho\left(\mathbb {E}\left(r_{\rm R N}\right), \Upsilon_{ v}, P_{\rm R N}\right)}-t_{2}\right).
\end{equation}

\section{NUMERICAL RESULTS}
To analyze the performance of SWITCH, numerical simulations are performed. The routing among SNs in simulations is based on the star graph $S_{4,2}$. The default parameter configuration is shown in the Table 1  \cite{b9} \cite{b34}.

\begin{table}[htpb]
\caption{Parameter settings}

\begin{tabular}{c|c}
\toprule
Parameter	                    &  Value       \\\hline
Transmitting power of RN    &  $P_{\rm R N}=30 d B m$             \\
Interference power          &  $P_{\rm I}=20 d B m$ \\
Noise power                 &  $\sigma_{z}^{2}=-55 d B m$ \\
Path loss coefficient       &  $\alpha=3$  \\
SINR target                 &  $\beta=0 d B$  \\
Time required to transmit data &  $t_{1}=50 \mu s$ \\
Transmission time of ACK or NACK signals  & $t_{2}=10 \mu s$\\
Radius                      &  $R=50m$  \\
Failure rate of SNs          &  $a_{1}=1.299 \times 10^{-7}$  \\
Failure rate of RNs          &  $a_{2}=5.299 \times 10^{-7}$  \\
Relationship coefficient between $\ell$ and ACC & $\varpi=1$  \\
Time interval               &   $T_{\rm s}=200$  \\
Maximum number of retransmissions per hop & $\mathrm{G}=8$  \\
Number of time slots in one time-interval &  $\tau=8$ \\

\bottomrule
\end{tabular}
\end{table}

Fig. 6 shows the success probability of data transmission as function of the power ratio $\mathcal{E}$ considering different interference intensities. Without loss of generality, the sector angle $\Phi$ is configured as $\Phi=\pi / 3$ in the following simulations. The SRN is located at $(-R,-R)$ and the DRN is located at $(R,R)$. When the power ratio is fixed, the success probability of data transmission decreases with the increase of the interference intensity. When the interference intensity is fixed, the success probability of data transmission increases with the increase of power ratio $\mathcal{E}$. When the power ratio $\varepsilon \geq 3$, the success probability of data transmission keeps a stable value. Based on the above result, the relationship between $P_{\rm R N}$ and $P_{\rm S N}$ is configured as $P_{\rm S N}=3 P_{\rm R N}$ to analyze the stable success probability of data transmission in the following simulations.
\begin{figure}[!h]
\centering
  \includegraphics[width=0.5\textwidth]{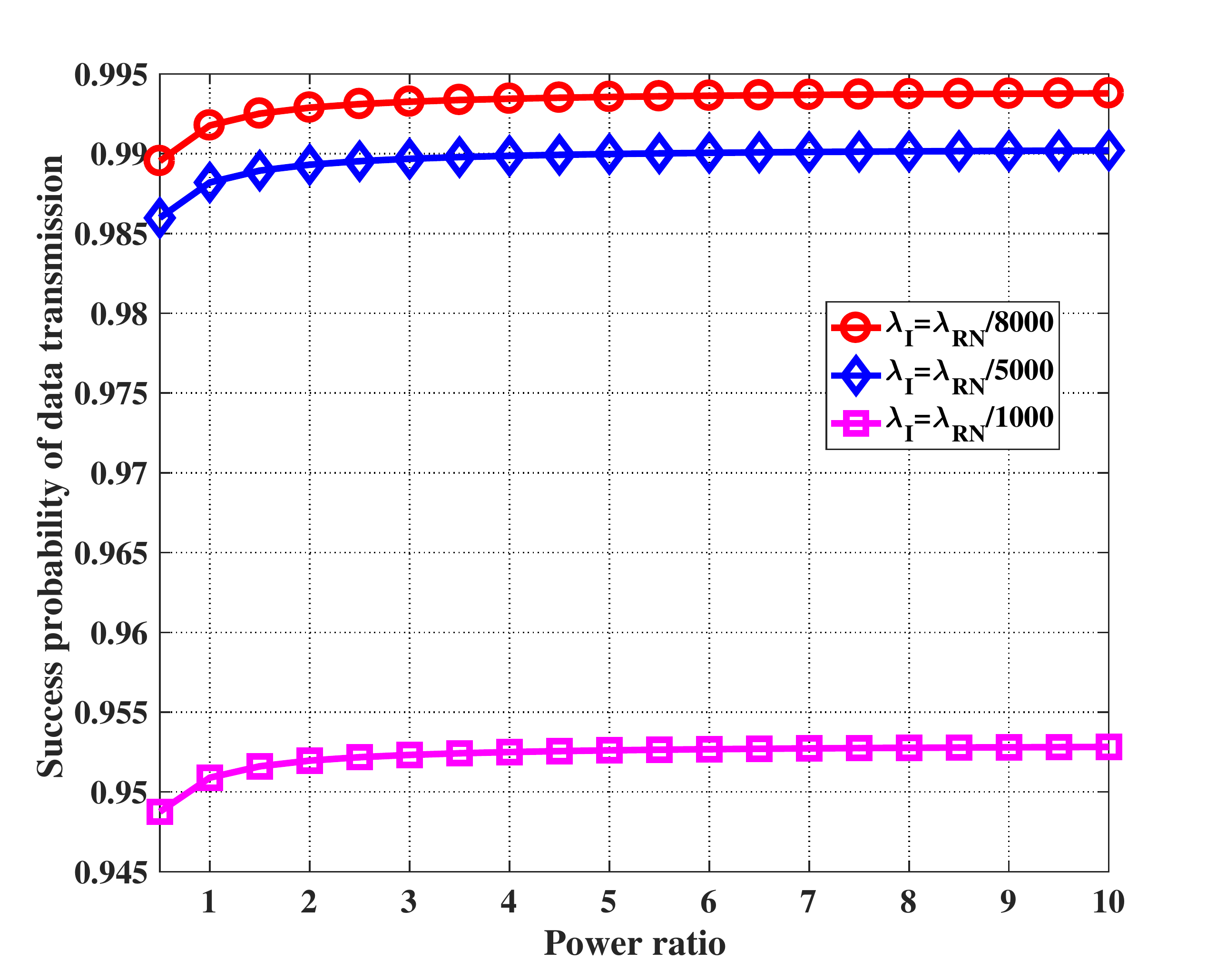}
\caption{ Success probability of data transmission as function of power ratio considering different interference intensities ($\lambda_{R N}=0.2$).}
\label{Fig6}       
\end{figure}

\begin{figure}[!h]
\centering
  \includegraphics[width=0.5\textwidth]{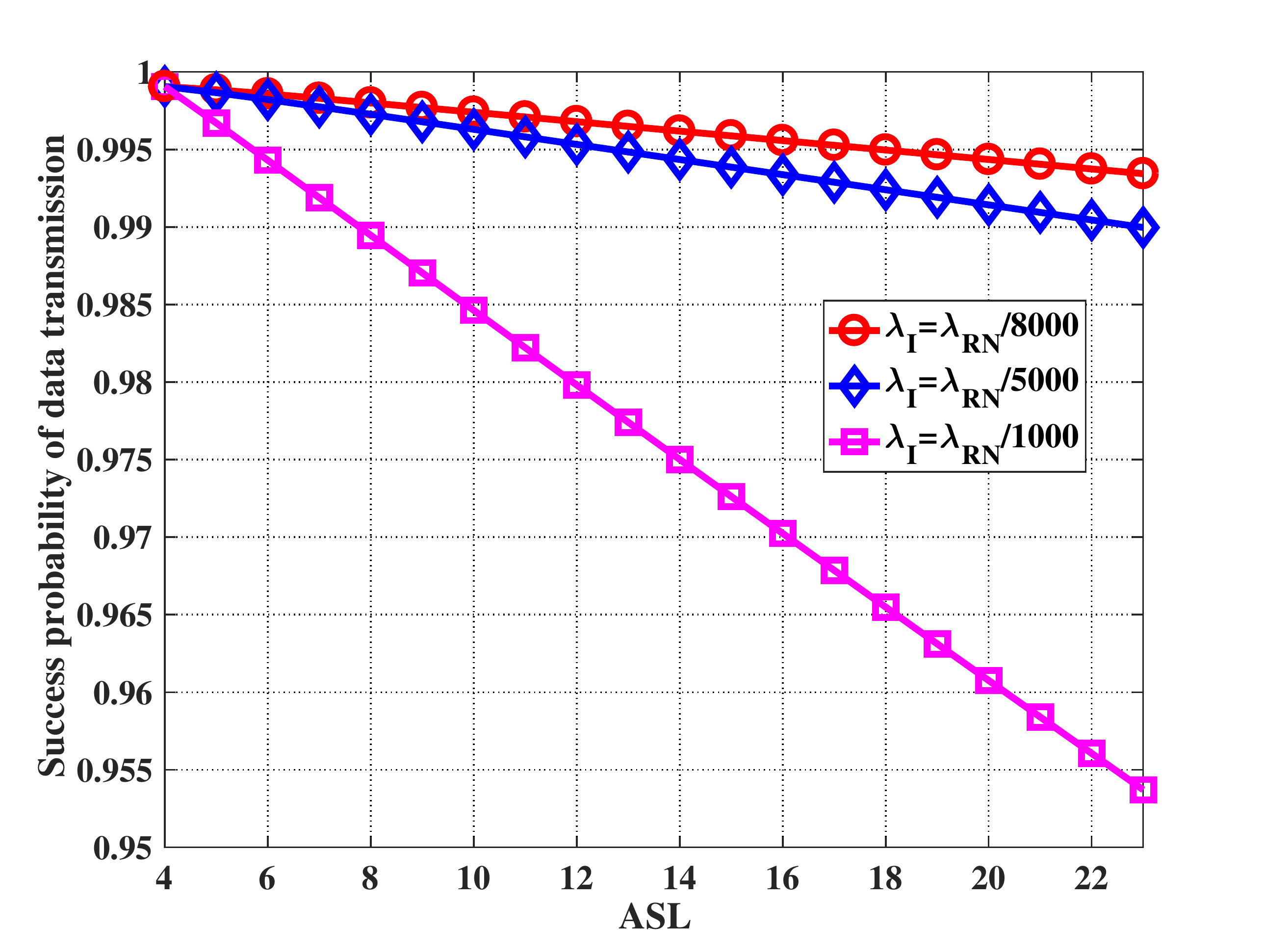}
\caption{Success probability of data transmission as function of ASL considering different interference intensities.}
\label{Fig7}       
\end{figure}

\begin{figure}[!h]
\centering
  \includegraphics[width=0.5\textwidth]{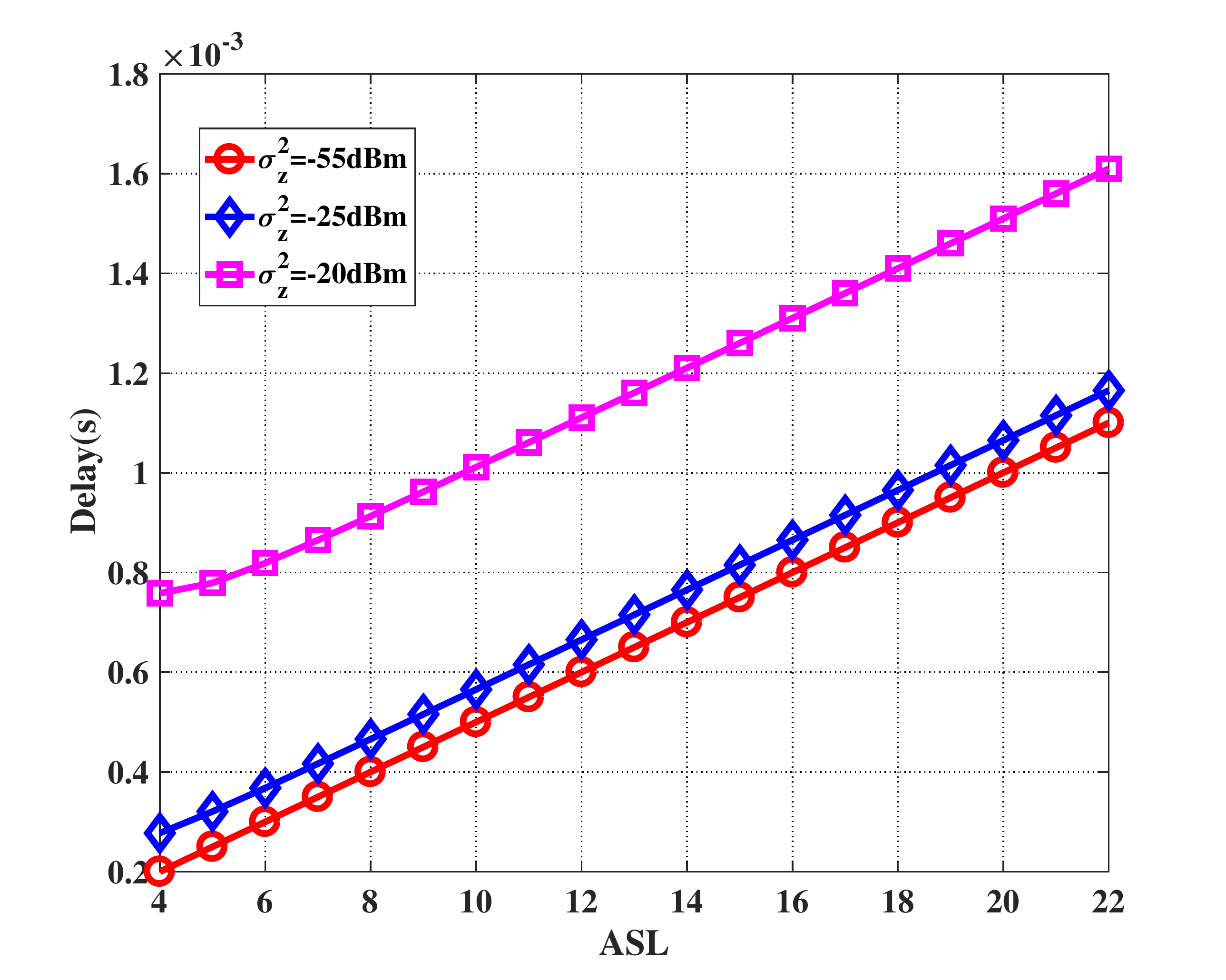}
\caption{Delay as function of ASL considering different noise power.}
\label{Fig8}       
\end{figure}

\begin{figure}[!h]
\centering
  \includegraphics[width=0.5\textwidth]{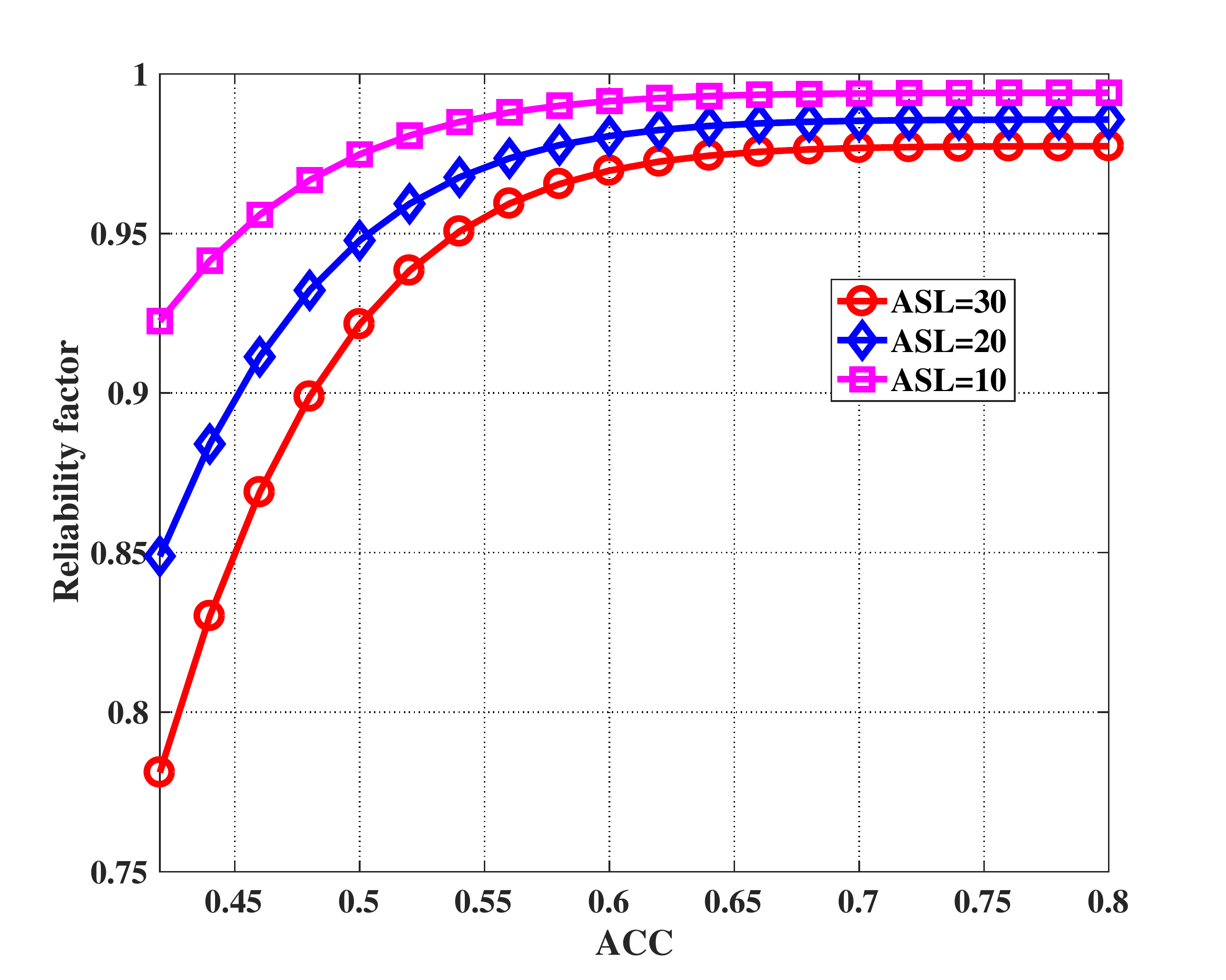}
\caption{Reliability factor as function of ACC considering different ASL.}
\label{Fig9}       
\end{figure}

Fig. 7 shows the success probability of data transmission as function of the ASL considering different interference intensities. When the relationship between the interference intensity and the density of RNs is fixed, the success probability of data transmission decreases with the increase of ASL. Since the increase of ASL leads to the increase of the interference intensity, which reduces the success probability of data transmission.

Fig. 8 shows the delay as function of the ASL considering different noise power in SWITCH. Without loss of generality, the interference intensity is configured as $\lambda_{\rm I}=\lambda_{\mathrm{RN}} / 8000$  in the following simulations. When the noise power is fixed, the delay grows linearly with the ASL as the processing delay increases with the increase of the ASL. Moreover, the delay increases with the increase of noise power when the ASL is fixed, due to the fact that the number of retransmissions increases with the growth of noise power.

Fig.9 shows the reliability factor as function of the ACC considering different ASLs in SWITCH. When the ASL is fixed, the reliability factor grows with the increase of ACC. When the ACC is larger than 0.6, the reliability factor approaches to 0.97, 0.98 and 0.99 when the ASL is configured as 30, 20 and 10, respectively. When the ACC is fixed, the reliability factor decreases with the increase of ASL.

Monte-Carlo simulations are conducted to compare SWITCH with NNR. Fig. 10 shows the reliability factor as function of the RN density $\lambda_{\rm R N}$ considering different sector angles $\Phi$. When $\Phi$ is fixed, the reliability factor decreases with the increase of RN density. Moreover, the reliability factor of SWITCH is higher than the reliability factor of NNR. When the RN density is fixed, the reliability factor decreases with the increase of $\Phi$.

Fig. 11 shows the delay as function of the RN density $\lambda_{\rm R N}$ considering different sector angles $\Phi$. When $\Phi$ is fixed, the delay increases with the increase of $\lambda_{\rm R N}$. Moreover, the delay of SWITCH is lower than the delay of NNR. When $\lambda_{\rm R N}$ is fixed, the delay increases with the increase of $\Phi$. Compared with the maximum delay of NNR, simulation results indicate that the maximum delay of SWITCH has been decreased by 50.6\%.
\begin{figure}[!h]
\centering
  \includegraphics[width=0.5\textwidth]{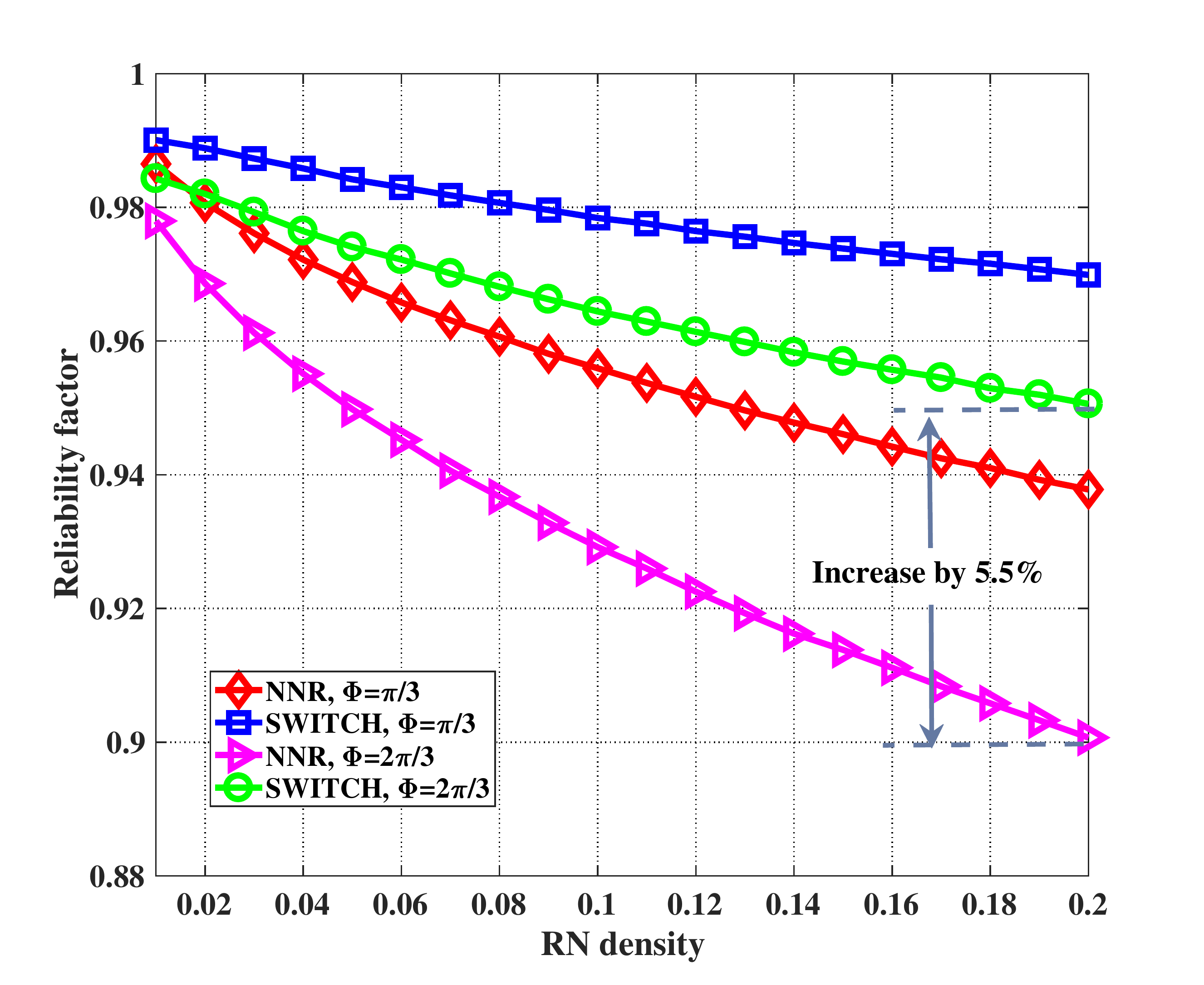}
\caption{Reliability factor as function of RN density considering different sector angles.}
\label{Fig10}       
\end{figure}
\begin{figure}[!h]
\centering
  \includegraphics[width=0.5\textwidth]{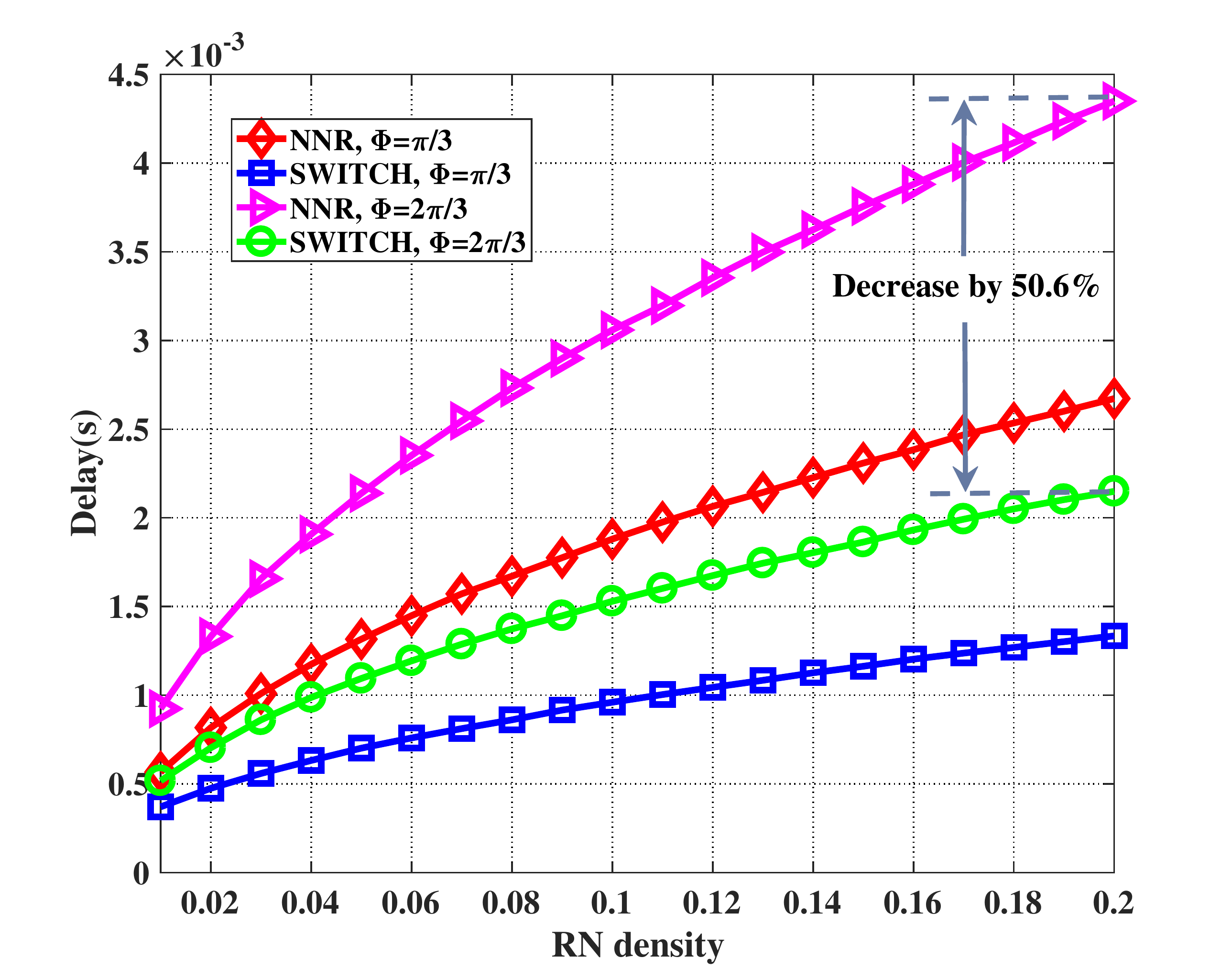}
\caption{Delay as function of RN density considering different sector angles.}
\label{Fig11}       
\end{figure}

\section{Conclusion}
In this paper, a new routing mechanism based on small-world characteristics was proposed for IoT networks. To build shortcuts in the routing strategies, a Cayley graph model was developed for IoT networks with different node transmission capabilities. Utilizing the metrics of ASL and ACC, the small-world characteristics of IoT networks based on Cayley graphs were analyzed and compared. Furthermore, the reliability and delay models were proposed for SWITCH based on ASL and ACC. Simulation results verified that SWITCH has small-world characteristics. Moreover, SWITCH has lower delay and higher reliability than that of NNR. Specifically, the maximum delay of SWITCH is reduced by 50.6\% compared with the maximum delay of NNR. As a consequence, SWITCH provides a low-delay and high reliable routing mechanism for data transmission in IoT networks. However, as the number of RNs scales up, the transmission data and consumed power of SNs have to be increased. How to balance the energy efficiency and transmission efficiency of SNs in IoT network with SWITCH solution remains a challenge. In the future, we will investigate how to improve the energy efficiency of IoT networks while ensuring low-delay and high-reliability.
\appendices
\section{Proof of Corollary 1}
In NNR, the number of neighbor RNs of the RN $v_{\rm R}$ is denoted as $b_{\rm R}$, the number of communication links existing among neighbor RNs of the RN $v_{\rm R}$ is denoted as $\delta_{\rm R}$. For the RN $v_{\rm R}$, $b_{\rm R}$ can be expressed as $b_{\rm R}=\lambda_{\mathrm{\rm RN}} \pi \left[\mathbb {E}\left(r_{\rm R N}\right)\right]^{2}=\pi^{2} / 2 \Phi$. The effects of boundary are not considered in this paper. The probability distribution of the distance $x_{\rm d}$ is derived by
\begin{equation}
F_{\rm v}\left(x_{\rm d}\right)=\frac{\pi x_{\rm d}^{2}}{\pi \left[\mathbb {E}\left(r_{\rm R N}\right)\right]^{2}}, \text { for } 0 \leq x_{\rm d} \leq \mathbb {E}\left(r_{\rm R N}\right).
\end{equation}
Then the probability density function of the distance $x_{\rm d}$ is expressed as
\begin{equation}
f_{\rm v}\left(x_{\rm d}\right)=\frac{2 x_{\rm d}}{\left[\mathbb {E}\left(r_{\rm R N}\right)\right]^{2}}, \text { for } 0 \leq x_{\rm d} \leq \mathbb {E}\left(r_{\rm R N}\right).
\end{equation}
The RN $v_{1}$ and the RN $v_{2}$ are two neighbor RNs of the RN $v_{\rm R}$. Configuring two disks with the same radius $\mathbb {E}\left(r_{\rm RN}\right)$ and centers  $v_{\rm R}$ and $v_{1}$, respectively, the overlap area of two disks is
\begin{equation}
\begin{array}{l}{\mathrm{S}_{\rm v_{1} \cap v_{R}}} \\ {=2 \arccos \left(\frac{x_{\rm d}}{2 \mathbb {E}\left(r_{\rm R N}\right)}\right) \left[\mathbb {E}\left(r_{\rm R N}\right)\right]^{2}-\frac{x_{\rm d} \sqrt{4 \left[\mathbb {E}\left(r_{\rm R N}\right)\right]^{2}-x_{\rm d}^{2}}}{2}}\end{array}.
\end{equation}
The probability that $v_{2}$ falls in the overlap area $\mathrm{S}_{\rm v_{1} \cap \rm v_{R}}$ is
\begin{equation}
\begin{aligned}
\operatorname{Pr}_{1}&=\frac{S_{\rm v_{1} \cap v_{\rm R}}}{\pi \left[\mathbb {E}\left(r_{\rm R N}\right)\right]^{2}}\\&=\frac{2 \arccos \left(\frac{x_{\rm d}}{2 \mathbb {E}\left(r_{\rm R N}\right)}\right)}{\pi}-\frac{x_{\rm d} \sqrt{4 \left[\mathbb {E}\left(r_{\rm R N}\right)\right]^{2}-x_{\rm d}^{2}}}{2 \pi \left[\mathbb {E}\left(r_{\rm R N}\right)\right]^{2}}
\end{aligned}.
\end{equation}
The probability that the distance between $v_{1}$ and $v_{2}$ is less than $\mathbb {E}\left(r_{\rm R N}\right)$, i.e., the probability that $v_{1}$ can communicate with $v_{2}$, is

\begin{equation}
\mathrm{Pr}_{2} = \int\limits_0^{E({r_{\rm RN}})} {{{\Pr }_1} \cdot {f_{\rm v}}({x_{\rm d}})d{x_{\rm d}}}.
\end{equation}
The number of communication links $\delta_{\rm R}$ is
\begin{equation}
\delta_{\rm R}=\begin{array}{c}{b_{\rm R}\choose 2}\end{array} \mathrm{Pr}_{2}.
\end{equation}
The clustering coefficient of the RN $v_{\rm R}$  is
\begin{equation}
C_{\rm R N}=\frac{2 \delta_{\rm R}}{b_{\rm R}\left(b_{\rm R}-1\right)}=\mathrm{Pr}_{2}.
\end{equation}
Hence, the ACC of NNR is
\begin{equation}
C_{\rm N N R}=\frac{1}{M_{\rm R}} \sum_{R N=1}^{M_{\rm R}} C_{\rm R N},
\end{equation}
where $M_{\rm R}=\lambda_{\rm R N}(2 R)^{2}$ is the number of RNs in the IoT network, the proof is completed.

\section{Proof of Theorem 1}
In SWITCH, the number of neighbor nodes including RNs and SNs of the SN $v_{\rm S}$ is denoted as $b_{\rm S}$. The number of communication links existing among neighbor nodes of the SN $v_{\rm S}$ is denoted as $\delta_{\rm S}$. $b_{\rm S}$ and $\delta_{\rm S}$ are expressed as
\begin{equation}
b_{\rm S}=b_{\rm R}+n-1,
\end{equation}
\begin{equation}
\delta_{\rm S}=\delta_{\rm R}+e_{\rm S}.
\end{equation}
The ACC of SWITCH is derived by
\begin{equation}
C_{\rm S W I T C H}=\frac{M_{\rm R} C_{\rm N N R}}{M_{\rm R}+N}+\frac{2 N \delta_{\rm S}}{b_{\rm S}\left(b_{\rm S}-1\right)\left(M_{\rm R}+N\right)}.
\end{equation}
The proof is completed.

\section*{Acknowledgment}

The authors would like to acknowledge the support from National Key R\&D Program of China (2016YFE0133000).

\begin{IEEEbiography}[{\includegraphics[width=1in,height=1.25in,clip,keepaspectratio]{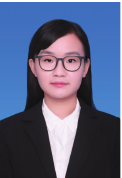}}]{Yuna Jiang}
received the B.E. degree in communications engineering from the China University of Mining and Technology, Xuzhou, China, in 2017. She is currently working toward the Ph.D. degree with the School of Electronic Information and Communications, Huazhong University of Science and Technology. Her research interests are Internet of Things and blockchain technology.
\end{IEEEbiography}

\vspace{-5 mm}

\begin{IEEEbiography}[{\includegraphics[width=1in,height=1.25in,clip,keepaspectratio]{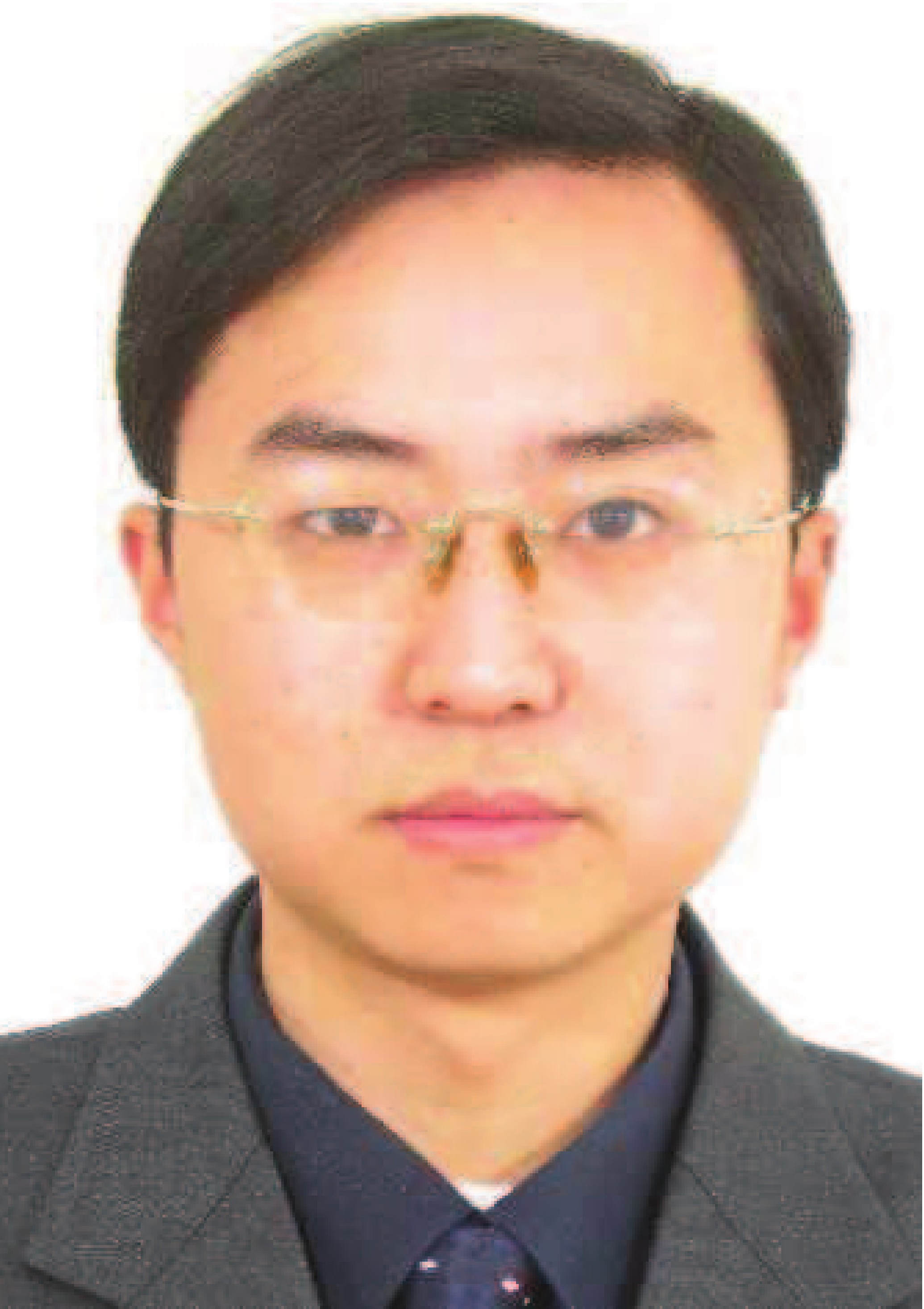}}]{Xiaohu Ge}
(M'09-SM'11) is currently a full Professor with the School of Electronic Information and Communications at Huazhong University of Science and Technology (HUST), China. He is an adjunct professor with the Faculty of Engineering and Information Technology at University of Technology Sydney (UTS), Australia. He received his PhD degree in Communication and Information Engineering from HUST in 2003. He has worked at HUST since Nov. 2005. Prior to that, he worked as a researcher at Ajou University (Korea) and Politecnico Di Torino (Italy) from Jan. 2004 to Oct. 2005. His research interests are in the area of mobile communications, traffic modeling in wireless networks, green communications, and interference modeling in wireless communications. He has published more than 200 papers in refereed journals and conference proceedings and has been granted about 25 patents in China. He received the Best Paper Awards from IEEE Globecom 2010. Dr. Ge served as the general Chair for the 2015 IEEE International Conference on Green Computing and Communications (IEEE GreenCom 2015). He serves as an associate editor for \textit{IEEE Wireless Communications}, \textit{IEEE Transactions on Vehicular Technology} and \textit{IEEE ACCESS}, etc.
\end{IEEEbiography}

\vspace{-5 mm}

\begin{IEEEbiography}[{\includegraphics[width=1in,height=1.25in,clip,keepaspectratio]{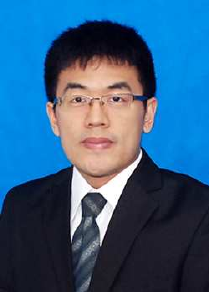}}]{Yi Zhong}
(S'12-M'15) received his B.S. and Ph.D. degree in Electronic Engineering from University of Science and Technology of China (USTC) in 2010 and 2015 respectively. From August to December 2012, he was a visiting student in Prof. Martin Haenggi's group at University of Notre Dame. From July to October 2013, he was an research intern with Qualcomm Incorporated, Corporate Research and Development, Beijing. From July 2015 to December 2016, he was a Postdoctoral Research Fellow with the Singapore University of Technology and Design (SUTD) in the Wireless Networks and Decision Systems (WNDS) Group led by Prof. Tony Q.S. Quek. Now, he is an assistant professor with School of Electronic Information and Communications, Huazhong University of Science and Technology, Wuhan, China. His main research interests include heterogeneous and femtocell-overlaid cellular networks, wireless ad hoc networks, stochastic geometry and point process theory.
\end{IEEEbiography}

\vspace{-5 mm}

\begin{IEEEbiography}[{\includegraphics[width=1in,height=1.25in,clip,keepaspectratio]{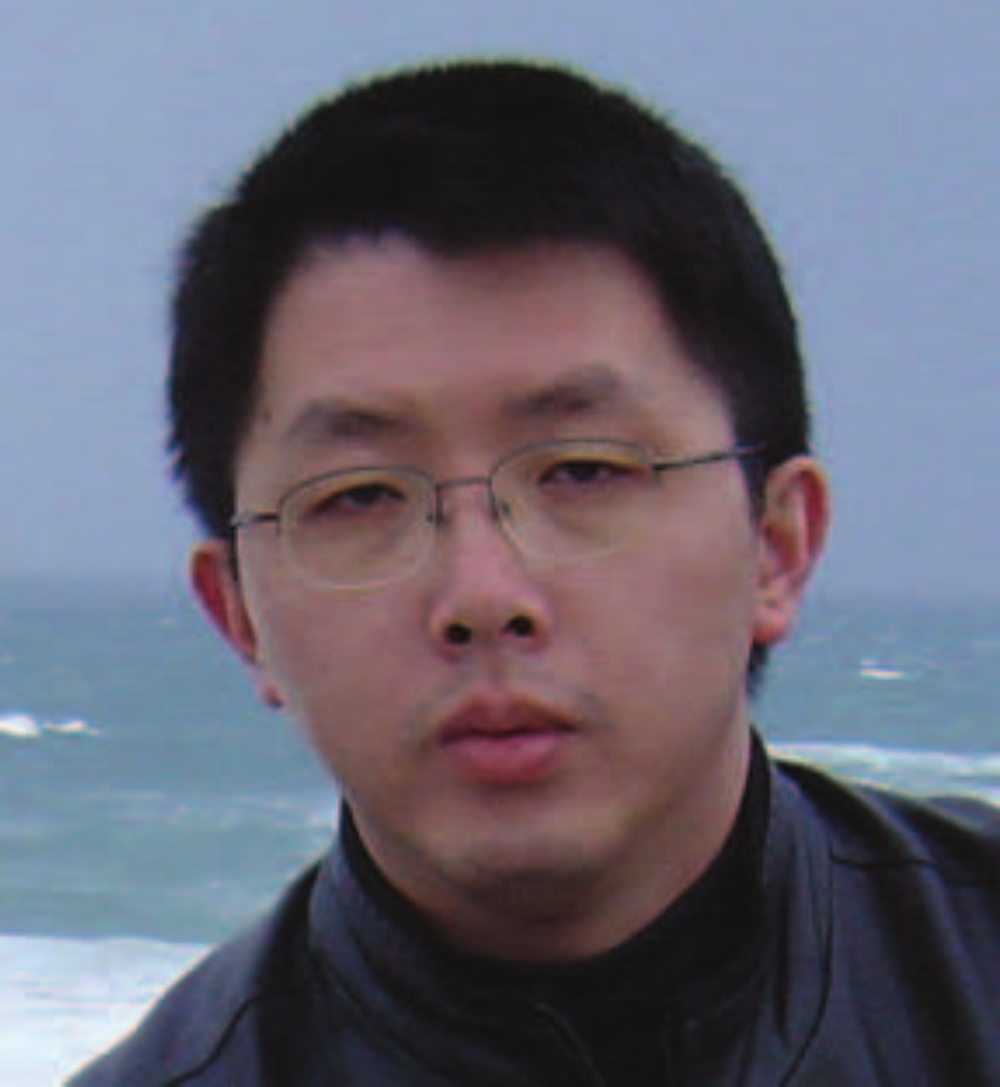}}]{Guoqiang~Mao}
(S'98-M'02-SM'08-F'18) joined the University of Technology Sydney in February 2014 as Professor of Wireless Networking and Director of Center for Real-time Information Networks. Before that, he was with the School of Electrical and Information Engineering, the University of Sydney. He has published about 200 papers in international conferences and journals, which have been cited more than 5000 times. He is an editor of the IEEE Transactions on Wireless Communications (since 2014), IEEE Transactions on Vehicular Technology (since 2010) and received \textquotedblleft Top Editor\textquotedblright{} award for outstanding contributions to the IEEE Transactions on Vehicular Technology in 2011, 2014 and 2015. He is a co-chair of IEEE Intelligent Transport Systems Society Technical Committee on Communication Networks. He has served as a chair, co-chair and TPC member in a large number of international conferences. He is a Fellow of IEEE and IET.

His research interest includes intelligent transport systems, applied graph theory and its applications in telecommunications, Internet of Things, wireless sensor networks, wireless localization techniques and network performance analysis.
\end{IEEEbiography}

\vspace{-5 mm}

\begin{IEEEbiography}[{\includegraphics[width=1in,height=1.25in,clip,keepaspectratio]{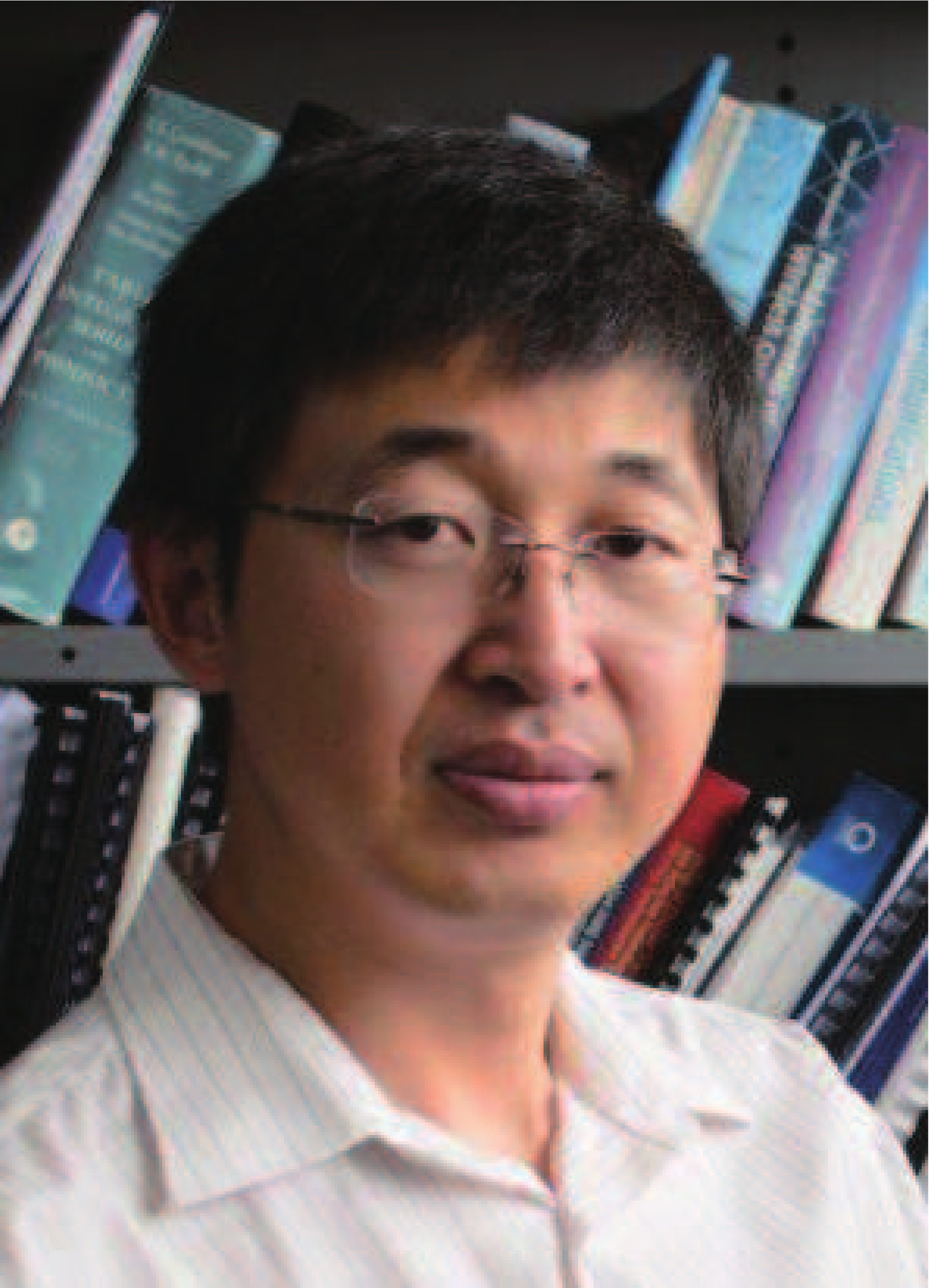}}]{Yonghui Li}
(M'04-SM'09-F'19) received his PhD degree in November 2002 from Beijing University of Aeronautics and Astronautics. From 1999 to 2003, he was affiliated with Linkair Communication Inc, where he held a position of project manager with responsibility for the design of physical layer solutions for the LAS-CDMA system. Since 2003, he has been with the Centre of Excellence in Telecommunications, the University of Sydney, Australia. He is now a Professor in School of Electrical and Information Engineering, University of Sydney. He is the recipient of the Australian Queen Elizabeth II Fellowship in 2008 and the Australian Future Fellowship in 2012.

His current research interests are in the area of wireless communications, with a particular focus on MIMO, millimeter wave communications, machine to machine communications, coding techniques and cooperative communications. He holds a number of patents granted and pending in these fields. He is now an editor for IEEE transactions on communications and IEEE transactions on vehicular technology. He was also the guest editor for IEEE JSAC Special issue on Millimeter Wave Communications for Future Mobile Networks. He received the best paper awards from IEEE International Conference on Communications (ICC) 2014, IEEE PIMRC 2017, and IEEE Wireless Days Conferences (WD) 2014. He is Fellow of IEEE.

\end{IEEEbiography}
\end{document}